\RequirePackage{fixltx2e}
\documentclass[12pt,a4paper]{iopart}
\pdfoutput=1
\usepackage[utf8]{inputenc}
\usepackage[unicode=true,bookmarksopen=true,colorlinks=true,urlcolor=blue,anchorcolor=blue,citecolor=blue,filecolor=blue,linkcolor=blue,menucolor=blue,pagecolor=blue,linktocpage=true,pdfa=true]{hyperref}

\usepackage{graphicx}

\makeatletter
\providecommand*\@nameundef[1]{\expandafter\let\csname #1\endcsname\@undefined}
\@nameundef{equation*}
\@nameundef{endequation*}
\makeatother
\usepackage{amsmath}
\usepackage{amssymb}
\usepackage{amsfonts}
\usepackage{mathtools}
\usepackage{lmodern}
\usepackage[T1]{fontenc}
\usepackage{bbm}
\DeclareMathAlphabet{\mathbfi}{OML}{cmm}{b}{it}

\usepackage{subcaption}

\let\originalleft\left
\let\originalright\right
\renewcommand{\left}{\mathopen{}\mathclose\bgroup\originalleft}
\renewcommand{\right}{\aftergroup\egroup\originalright}

\makeatletter
\newenvironment{equations}[1][]{\subequations\ifx\relax#1\relax\else\label{#1}\fi\align\ignorespaces}{\endalign\ignorespacesafterend\endsubequations}
\def\@spliteq#1{\begin{equation}\begin{split}#1\end{split}\end{equation}}
\def\splitequation{\collect@body\@spliteq}

\makeatother

\renewcommand{\vec}[1]{{\ifnum9<1#1\mathbf{#1}\else\ifcat\noexpand#1\relax\boldsymbol{#1}\else\mathbfi{#1}\fi\fi}}
\newcommand{\mathe}{\mathrm{e}}
\newcommand{\mathi}{\mathrm{i}}
\let\oldre\Re
\let\oldim\Im
\renewcommand{\Re}{\oldre\mathfrak{e}\,}
\renewcommand{\Im}{\oldim\mathfrak{m}\,}
\newcommand{\total}{\mathop{}\!\mathrm{d}}

\newcommand{\eqend}[1]{\,#1}
\newcommand{\bigo}[1]{\mathcal{O}\left({#1}\right)}

\newcommand{\expect}[1]{\left\langle{#1}\right\rangle}

\usepackage[numbers,sort&compress]{natbib}
\allowdisplaybreaks

\begin{document}

\title[Scalar two-point function at fixed geodesic distance]{One-loop quantum gravitational corrections to the scalar two-point function at fixed geodesic distance}

\author{Markus B. Fröb}
\address{Department of Mathematics, University of York, Heslington, York, YO10 5DD, United Kingdom}
\ead{mbf503@york.ac.uk}

\begin{abstract}
We study a proposal for gauge-invariant correlation functions in perturbative quantum gravity, which are obtained by fixing the geodesic distance between points in the fluctuating geometry. These correlation functions are non-local and strongly divergent, and we show how to renormalise them by performing a ``wave function renormalisation'' of the geodesic embedding coordinates. The result is finite and gauge-independent, but displays unusual features such as double logarithms at one-loop order.

\noindent\textit{Keywords}: perturbative quantum gravity, geodesic distance, renormalisation
\end{abstract}

\pacs{04.60.-m, 04.62.+v, 11.10.Gh, 11.15.-q}
\submitto{CQG}

\maketitle

\section{Introduction}
\label{sec_introduction}

While a full theory of quantum gravity is still elusive, much can be learned from studying quantum gravity in the sense of an effective field theory~\cite{burgess2004}, by considering metric fluctuations around a given classical background. Even though this theory is not power-counting renormalisable, one can make predictions which are valid at scales well below the Planck scale, which includes the only observational evidence for quantum gravity to date, the cosmic microwave background~\cite{planck2015a,planck2015b,planck2015c}. In particular, unambiguous predictions can be made for quantum corrections to the Newtonian and other gravitational potentials of point particles due to matter loops in flat space~\cite{schwinger1968,radkowski1970,duff1974,duffliu2000,satzmazzitellialvarez2005,parkwoodard2010,marunovicprokopec2011,marunovicprokopec2012,burnspilaftsis2015,froeb2016} and de~Sitter space~\cite{wangwoodard2015b,parkprokopecwoodard2016,froebverdaguer2016a,froebverdaguer2017}, for quantum corrections to the time delay in various scenarios~\cite{khavkine2012,bongakhavkine2014,battistaetal2017}, corrections to Solar system dynamics~\cite{battistaesposito2014,battistaetal2015}, corrections to geodesic deviation~\cite{borgmanford2004,dragopinamonti2014}, etc. However, the situation changes drastically when one wants to include the effects of graviton loops (calling any metric fluctuation ``graviton'' for short, and not only the transverse traceless part), for example to calculate quantum gravitational corrections in electrodynamics~\cite{bjerrumbohr2002,donoghueetal2002,fordhertzbergkarouby2016}. In perturbative quantum gravity, diffeomorphism invariance translates into a gauge symmetry for the graviton, and while one can find local gauge-invariant gravitational observables at linear order (e.g., the linearised Riemann tensor for a flat-space background or the linearised Weyl tensor for conformally flat backgrounds), in general this is impossible (see, e.g., refs.~\cite{torre1993,giddingsmarolfhartle2006,khavkine2015} for proofs with various levels of mathematical sophistication). In flat space one can nevertheless calculate corrections to, say, the Newtonian potential by reconstructing a scattering potential which gives the same S-matrix element as the perturbative calculation including graviton loops~\cite{donoghue1994a,muzinichvokos1995,hamberliu1995,akhundovbelluccishiekh1997,kirilinkhriplovich2002,donoghueetal2002,khriplovichkirilin2003,bjerrumbohrdonoghueholstein2003a,bjerrumbohrdonoghueholstein2003b,holsteinross2008}. Since the S-matrix is gauge-independent, the resulting potential is invariant as well if one restricts, as usual, to diffeomorphisms which decay sufficiently fast at infinity. This approach fails for spacetimes for which no S-matrix can be defined; moreover, one would expect that a potential which is measured locally, in a region of finite size, can be obtained from a local (or at least almost local) observable in the quantum theory, without resorting to non-local observables which are defined at infinity.

There are various approaches to the problem of defining such almost local observables in a (at least perturbative) theory of quantum gravity, among which predominant ones are the following:
\begin{itemize}
\item ``Dressing'' of the bare field operators and states with a graviton cloud~\cite{waresaotomeakhoury2013,donnellygiddings2015,donnellygiddings2016}, either extended or string-like. This is analogous to the construction of dressed operators and states in QED~\cite{kibble1968,kulishfaddeev1970,steinmann1984,baganlavellemcmullan2000a,baganlavellemcmullan2000b,mitraratabolesharatchandra2006}, going back to the work of Dirac~\cite{dirac1955}.
\item Relational observables, considering the value of a field at the point where another field has a given value~\cite{giddingsmarolfhartle2006,khavkine2015,marolf2015,brunettifredenhagenrejzner2016}. This goes back to early work by G{\'e}h{\'e}niau and Deveber~\cite{geheniaudebever1956a,geheniau1956,debever1956a,debever1956b,geheniaudebever1956b}, extended by Komar and Bergmann~\cite{komar1958,bergmannkomar1960,bergmann1961}; see ref.~\cite{tambornino2012} for a recent review. Generally, these observables use scalars constructed from various background fields as coordinates, and thus require a sufficiently generic background spacetime, but an extension of the concept to highly symmetric spacetimes has recently been achieved~\cite{brunettietal2016}.
\item Defining the distance between points in correlation functions using perturbed geodesics \cite{tsamiswoodard1992,modanese1994,urakawatanaka2009,urakawatanaka2010}, similar to Mandelstam's path-dependent reformulation of field theory~\cite{mandelstam1962,mandelstam1968,teitelboim1993}. This approach has been pioneered by Woodard~\cite{woodard_thesis}, and has also been used in numerical studies~\cite{hamber1994,ambjornanagnostopoulos1997}. We adopt it in the following.
\end{itemize}

To define an $n$-point scalar correlation function at fixed geodesic distance, one fixes a starting point $x^\mu$, $n$ starting directions $v_{(n)}^\mu$ and $n$ geodesic distances $\ell_{(n)}$. One has then to solve the geodesic equation
\begin{equation}
\label{introduction_geodesiceq}
\ddot \chi_{(n)}^\mu(\tau) + \Gamma^\mu_{\alpha\beta}(\chi_{(n)}(\tau)) \dot \chi_{(n)}^\alpha(\tau) \dot \chi_{(n)}^\beta(\tau) = 0 \eqend{,}
\end{equation}
where a dot denotes a derivative with respect to the affine parameter $\tau$, with the boundary conditions $\chi_{(n)}^\mu(0) = x^\mu$, $\dot \chi_{(n)}^\mu(0) = v_{(n)}^\mu$, and calculate the geodesic lengths
\begin{equation}
s_{(n)}(\tau) = \int_0^\tau \sqrt{ g_{\mu\nu}\left( \chi(\tau') \right) \dot \chi_{(n)}^\mu(\tau') \dot \chi_{(n)}^\nu(\tau') } \total \tau' \eqend{.}
\end{equation}
Let us assume w.l.o.g. that the affine parameter is normalised such that $s_{(n)}(1) = \ell_{(n)}$; the sought correlation function is then given by
\begin{equation}
\expect{ \phi(\chi_{(1)}(1)) \cdots \phi(\chi_{(n)}(1)) } \eqend{.}
\end{equation}
In addition to the above, one has to parallel transport tensor or spinor fields along the geodesic back to the origin to compare them in a common Lorentz frame~\cite{woodard_thesis,tsamiswoodard1992,modanese1994}. One can then calculate this expectation value expanding both the action and the geodesic perturbatively around a background. In the present work, we focus on one-loop corrections to the two-point function around flat space, and calculate
\begin{equation}
\label{introduction_2pf}
\expect{ \phi(\chi(0)) \phi(\chi(1)) }
\end{equation}
to order $\kappa^2 = 16 \pi G_\text{N}$, where $G_\text{N}$ is Newton's constant. The main open problem in a perturbative calculation of such correlation functions is their renormalisation. Solving the geodesic equation~\eqref{introduction_geodesiceq} perturbatively with the given boundary conditions, one obtains terms where the graviton is integrated over the background geodesic, a one-dimensional submanifold. In field theory one deals with distribution-valued operators, and their restriction to submanifolds is not well defined in general, leading to new divergences. While these divergences can be regularised, e.g., in dimensional regularisation, they can not be renormalised using the usual bulk action counterterms. Previous calculations have either left this problem open~\cite{woodard_thesis,tsamiswoodard1992}, or imposed an ad-hoc small distance cut-off on the integration over the background geodesic~\cite{modanese1992,modanese1994}. Borrowing ideas from the renormalisation of Wilson loops in non-Abelian gauge theories~\cite{gervaisneveu1980,dotsenkovergeles1980,brandtnerisato1981} and post-Newtonian point particle dynamics in General Relativity~\cite{blanchetdamourespositofarese2004,blanchetetal2005,jaranowskischaefer2013,bernardetal2015}, we show that these additional divergences can be renormalised by a ``wave function renormalisation'' of the geodesic itself, of the form $\chi^\mu(\tau) \to Z_\chi \chi^\mu(\tau)$, at least to order $\kappa^2$.

The article is organised as follows: section~\ref{sec_calculation} shows the calculation of the correlation function~\eqref{introduction_2pf} in detail, separating purely field-theoretic contributions from purely geodesic corrections and mixed ones, and performing renormalisation. Section~\ref{sec_results} gives the final result, and discusses its renormalisation group scaling and the gauge (in-)dependence, and section~\ref{sec_discussion} discusses its significance. Some technical computations are relegated to the appendices. We use the `+++' convention of ref.~\cite{mtw_book}, and set $c = \hbar = 1$ and $\kappa^2 = 16 \pi G_\text{N}$.

\section{Calculation}
\label{sec_calculation}

\subsection{Preliminaries}
\label{sec_calculation_prelim}

For simplicity, we restrict to the case of a massless scalar, but include arbitrary coupling of the scalar to curvature (parametrised by the constant $\xi$, with $\xi = (n-2)/[4(n-1)]$ corresponding to conformal coupling). We then consider the standard Einstein-Klein Gordon action
\begin{equation}
\label{calculation_prelim_action_ekg}
S = \frac{1}{\kappa^2} \int R \sqrt{-g} \total^n x - \frac{1}{2} \int \left( g^{\mu\nu} \partial_\mu \phi \partial_\nu \phi + \xi R \phi^2 \right) \sqrt{-g} \total^n x \eqend{.}
\end{equation}
Since massless tadpoles vanish in dimensional regularisation in flat space, and mass is multiplicatively renormalised, we need neither a counterterm for the cosmological constant nor for the scalar mass (and to lowest non-trivial order in $\kappa$ also no wave function renormalisation is needed), but we have to include a higher-order counterterm of the form
\begin{equation}
\label{calculation_prelim_action_ct}
S_\text{CT} = \delta_{(\partial^2 \phi)^2} \int \left( \nabla^2 \phi \right)^2 \sqrt{-g} \total^n x \eqend{.}
\end{equation}
We also need to include gauge-fixing and ghost terms, but since we only consider external scalars and work to lowest non-trivial order in $\kappa$, the ghost terms are irrelevant (as well as the auxiliary field). The most general linear gauge-fixing action for pure gravity, depending on two parameters $\alpha$ and $\beta$, is then given by
\begin{equation}
S_\text{GF} = - \frac{1}{2 \alpha} \int \left[ \partial^\nu h_{\mu\nu} - \left( 1 + \frac{1}{\beta} \right) \partial_\mu h \right] \left[ \partial_\rho h^{\mu\rho} - \left( 1 + \frac{1}{\beta} \right) \partial^\mu h \right] \total^n x \eqend{.}
\end{equation}

We perturbatively expand the action~\eqref{calculation_prelim_action_ekg} around a flat background
\begin{equation}
g_{\mu\nu} = \eta_{\mu\nu} + \kappa h_{\mu\nu}
\end{equation}
to second order in $h_{\mu\nu}$, which can be done using the expansions from~\ref{appendix_metric}. We obtain
\begin{equation}
S + S_\text{GF} = S_0 + \kappa S_1^{h\phi\phi} + \kappa S_1^{h^3} + \kappa^2 S_2 + \bigo{\kappa^3}
\end{equation}
with the free action
\begin{splitequation}
\label{calculation_prelim_s0}
S_0 = \frac{1}{2} \int h^{\mu\nu} P_{\mu\nu\rho\sigma} h^{\rho\sigma} \total^n x + \frac{1}{2} \int \phi \partial^2 \phi \total^n x
\end{splitequation}
and the first order graviton--scalar interaction
\begin{equation}
S_1^{h\phi\phi} = \frac{1}{2} \int \left( \tau_{\mu\nu\rho\sigma} \partial^\rho \phi \partial^\sigma \phi - \xi S_{\mu\nu} \phi^2 \right) h^{\mu\nu} \total^n x \eqend{,}
\end{equation}
where we defined the tensor
\begin{equation}
\label{calculation_prelim_tau}
\tau_{\mu\nu\rho\sigma} \equiv \eta_{\rho(\mu} \eta_{\nu)\sigma} - \frac{1}{2} \eta_{\mu\nu} \eta_{\rho\sigma}
\end{equation}
and the symmetric differential operators
\begin{equations}
S_{\mu\nu} &\equiv \partial_\mu \partial_\nu - \eta_{\mu\nu} \partial^2 \eqend{,} \\
\begin{split}
P_{\mu\nu\rho\sigma} &\equiv \frac{1}{2} \eta_{\mu(\rho} \eta_{\sigma)\nu} \partial^2 - \left( 1 - \frac{1}{\alpha} \right) \partial_{(\mu} \eta_{\nu)(\rho} \partial_{\sigma)} \\
&\quad+ \left( \frac{1}{2} - \frac{1 + \beta}{\alpha \beta} \right) \left( \eta_{\mu\nu} \partial_\rho \partial_\sigma + \eta_{\rho\sigma} \partial_\mu \partial_\nu \right) - \left( \frac{1}{2} - \frac{(1+\beta)^2}{\alpha \beta^2} \right) \eta_{\mu\nu} \eta_{\rho\sigma} \partial^2 \eqend{.}
\end{split}
\end{equations}
Since massless tadpoles, which arise from $S_2$, correspond to scaleless integrals in momentum space, they vanish in dimensional regularisation~\cite{leibbrandt1975}, and we do not need the explicit expression for $S_2$. Similarly, $S_1^{h^3}$ gives the vertex corresponding to a 3-graviton interaction, which is not needed in our case.

The propagators are as usual obtained by inverting the differential operators appearing in the free action~\eqref{calculation_prelim_s0}, and we obtain
\begin{equation}
G_0(x,y) = \int \tilde{G}_0(p) \mathe^{\mathi p (x-y)} \frac{\total^n p}{(2\pi)^n} = - \int \frac{1}{p^2 - \mathi 0} \mathe^{\mathi p (x-y)} \frac{\total^n p}{(2\pi)^n}
\end{equation}
satisfying
\begin{equation}
\label{calculation_prelim_eom_scalar}
\partial^2 G_0(x,y) = \delta^n(x-y)
\end{equation}
for the scalar field and
\begin{splitequation}
G_{\mu\nu\rho\sigma}(x,y) &= \left( 2 \eta_{\mu(\rho} \eta_{\sigma)\nu} - \frac{2}{n-2} \eta_{\mu\nu} \eta_{\rho\sigma} \right) G_0(x,y) + 4 (\alpha-1) \frac{\partial_{(\mu} \eta_{\nu)(\rho} \partial_{\sigma)}}{\partial^2} G_0(x,y) \\
&\quad+ \frac{2}{n-2} (2+\beta) \left( \eta_{\mu\nu} \frac{\partial_\rho \partial_\sigma}{\partial^2} + \eta_{\rho\sigma} \frac{\partial_\mu \partial_\nu}{\partial^2} \right) G_0(x,y) \\
&\quad- (2+\beta) \left[ \frac{n}{n-2} (2+\beta) + (\alpha-1) (2-\beta) \right] \frac{\partial_\mu \partial_\nu \partial_\rho \partial_\sigma}{\left( \partial^2 \right)^2} G_0(x,y)
\end{splitequation}
satisfying
\begin{equation}
P^{\mu\nu\alpha\beta} G_{\alpha\beta\rho\sigma}(x,y) = \delta^{(\mu}_\rho \delta^{\nu)}_\sigma \delta^n(x-y)
\end{equation}
for the graviton field. For later use, we note that the explicit expression in coordinate space for $G_0$ reads
\begin{equation}
\label{calculation_prelim_g0}
G_0(x,y) = - \mathi \frac{c_n}{[(x-y)^2]^\frac{n-2}{2}}
\end{equation}
with the constant
\begin{equation}
\label{calculation_prelim_cn_def}
c_n \equiv \frac{\Gamma\left( \frac{n-2}{2} \right)}{4 \pi^\frac{n}{2}} = \frac{1}{4 \pi^2} \left[ 1 - \frac{n-4}{2} \left( \gamma + \ln \pi \right) + \frac{(n-4)^2}{8} \left[ \frac{\pi^2}{6} + \left( \gamma + \ln \pi \right)^2 \right] \right] + \bigo{(n-4)^3} \eqend{,}
\end{equation}
where $\gamma$ is the Euler--Mascheroni constant. Defining the five tensors
\begin{equations}[calculation_prelim_gravprop_tensors]
T^{(1)}_{\mu\nu\rho\sigma}(k) &= 2 \eta_{\mu(\rho} \eta_{\sigma)\nu} \eqend{,} \\
T^{(2)}_{\mu\nu\rho\sigma}(k) &= \eta_{\mu\nu} \eta_{\rho\sigma} \eqend{,} \\
T^{(3)}_{\mu\nu\rho\sigma}(k) &= \eta_{\mu\nu} \frac{k_\rho k_\sigma}{k^2} + \eta_{\rho\sigma} \frac{k_\mu k_\nu}{k^2} \eqend{,} \\
T^{(4)}_{\mu\nu\rho\sigma}(k) &= 4 \frac{k_{(\mu} \eta_{\nu)(\rho} k_{\sigma)}}{k^2} \eqend{,} \\
T^{(5)}_{\mu\nu\rho\sigma}(k) &= \frac{k_\mu k_\nu k_\rho k_\sigma}{k^4} \eqend{,}
\end{equations}
the momentum-space graviton propagator can be written as
\begin{equation}
\label{calculation_prelim_gravprop}
\tilde{G}_{\mu\nu\rho\sigma}(k) = - \sum_{i=1}^5 g_i T^{(i)}_{\mu\nu\rho\sigma}(k) \frac{1}{k^2 - \mathi 0}
\end{equation}
with the coefficients
\begin{splitequation}
g_1 &= 1 \eqend{,} \quad g_2 = - \frac{2}{n-2} \eqend{,} \quad g_3 = \frac{2}{n-2} (2+\beta) \eqend{,} \\
g_4 &= \alpha-1 \eqend{,} \quad g_5 = - (2+\beta) \left[ \frac{n}{n-2} (2+\beta) + (\alpha-1) (2-\beta) \right] \eqend{.}
\end{splitequation}
The analogue of Feynman gauge in electromagnetism is achieved for $\alpha = 1$ and $\beta = -2$, while a one-parameter family of gauges analogous to Landau gauge is obtained for $\alpha = 0$. However, to explicitly show the gauge dependence of our result we will keep $\alpha$ and $\beta$ arbitrary. We also calculate the trace
\begin{splitequation}
\label{calculation_prelim_trace}
\eta^{\mu\nu} G_{\mu\nu\rho\sigma}(x,y) &= \frac{2}{n-2} \beta \eta_{\rho\sigma} G_0(x,y) \\
&\quad+ \beta \left[ (\alpha-1) \beta - \frac{n}{n-2} (2+\beta) \right] \frac{\partial_\rho \partial_\sigma}{\partial^2} G_0(x,y) \eqend{,}
\end{splitequation}
which vanishes for $\beta = 0$.

Since we want to calculate the two-point function at fixed geodesic distance, we also have to perturbatively expand the equation for the geodesic and its length. Consider thus a geodesic $\chi^\mu(\tau)$ which fulfils the geodesic equation
\begin{equation}
\label{calculation_prelim_geodesic_eqn}
\ddot \chi^\mu(\tau) + \Gamma^\mu_{\alpha\beta}\left( \chi(\tau) \right) \dot \chi^\alpha(\tau) \dot \chi^\beta(\tau) = 0
\end{equation}
and the initial conditions
\begin{equation}
\label{calculation_prelim_geodesic_bdy1}
\chi^\mu(0) = x^\mu \eqend{,} \quad \dot \chi^\mu(0) = v^\mu
\end{equation}
for some vector $v^\mu$. The geodesic length $\ell$ is given by
\begin{equation}
\ell = \int_0^1 \sqrt{ g_{\mu\nu}\left( \chi(\tau) \right) \dot \chi^\mu(\tau) \dot \chi^\nu(\tau) } \total \tau \eqend{,}
\end{equation}
and since using the geodesic equation we have
\begin{splitequation}
\frac{\total}{\total \tau} \big[ g_{\mu\nu}\left( \chi(\tau) \right) \dot \chi^\mu(\tau) \dot \chi^\nu(\tau) \big] &= \big[ \partial_\alpha g_{\mu\nu}\left( \chi(\tau) \right) \dot \chi^\alpha(\tau) \dot \chi^\mu(\tau) + 2 g_{\mu\nu}\left( \chi(\tau) \right) \ddot \chi^\mu(\tau) \big] \dot \chi^\nu(\tau) \\
&= \big[ \partial_\alpha g_{\beta\nu}\left( \chi(\tau) \right) - 2 g_{\mu\nu}\left( \chi(\tau) \right) \Gamma^\mu_{\alpha\beta}\left( \chi(\tau) \right) \big] \dot \chi^\alpha(\tau) \dot \chi^\beta(\tau) \dot \chi^\nu(\tau) \\
&= 0 \eqend{,}
\end{splitequation}
it follows that
\begin{equation}
\label{calculation_prelim_mu2}
\ell^2 = g_{\mu\nu}\left( \chi(0) \right) \dot \chi^\mu(0) \dot \chi^\nu(0) = g_{\mu\nu}(x) v^\mu v^\nu \eqend{.}
\end{equation}
To lowest order, the geodesic equation~\eqref{calculation_prelim_geodesic_eqn} with boundary conditions~\eqref{calculation_prelim_geodesic_bdy1} is solved by
\begin{equation}
\label{calculation_prelim_geodesic_order0}
\chi^\mu(\tau) = x^\mu + v^\mu \tau \eqend{,}
\end{equation}
and taking $v^\mu = - (x-y)^\mu$ the endpoint of the geodesic is given by $\chi^\mu(1) = y^\mu$. At higher orders in $\kappa$, the explicit metric appearing in equation~\eqref{calculation_prelim_mu2} shows that $v^\mu$ must be corrected if we want to fix $\ell$. This is most easily done by introducing an $n$-bein $e_\mu{}^a$ which satisfies the relation
\begin{equation}
g_{\mu\nu} = e_\mu{}^a e_\nu{}^b \eta_{ab}
\end{equation}
with a flat metric $\eta_{ab}$. It then follows that
\begin{equation}
\ell^2 = \eta_{ab} e_\mu{}^a e_\nu{}^b v^\mu v^\nu = \eta_{ab} v^a v^b \eqend{,}
\end{equation}
such that fixing the geodesic distance means fixing the bein components $v^a$, i.e., the initial condition for the geodesic must be taken as
\begin{equation}
\label{calculation_prelim_geodesic_bdy2}
\dot \chi^\mu(0) = g^{\mu\nu} e_\nu{}^a v^b \eta_{ab}
\end{equation}
with constant $v^b$~\cite{woodard_thesis}. In perturbation theory, it is extremely convenient to fix the local Lorentz symmetry of the $n$-bein to symmetric gauge, which completely decouples the corresponding ghosts and permits to express the perturbed $n$-bein using the metric perturbation $h_{\mu\nu}$~\cite{vannieuwenhuizen1981,woodard_thesis,woodard1984}, given by equation~\eqref{appendix_metric_bein_expansion}.

Let us now derive the geodesic corrections to first order in $\kappa$, writing $\chi^\mu(\tau) = \chi_0^\mu(\tau) + \kappa \chi_1^\mu(\tau) + \bigo{\kappa^2}$. The boundary condition~\eqref{calculation_prelim_geodesic_bdy2} reads
\begin{equation}
\dot \chi^\mu(0) = v^\mu - \frac{1}{2} \kappa h^\mu_\nu v^\nu + \bigo{\kappa^2} \eqend{,}
\end{equation}
such that
\begin{equation}
\label{calculation_prelim_geodesic_bdy_firstorder}
\dot \chi_1^\mu(0) = - \frac{1}{2} h^\mu_\nu(x) v^\nu \eqend{.}
\end{equation}
Since $\chi^\mu(0) = x^\mu$ is fixed once and for all, there are no corrections to the starting point, such that $\chi_1^\mu(0) = 0$. Expanding now the geodesic equation~\eqref{calculation_prelim_geodesic_eqn} to first order and using that the lowest-order geodesic is given by equation~\eqref{calculation_prelim_geodesic_order0}, we get
\begin{equation}
\ddot \chi_1^\mu(\tau) = v^\alpha v^\beta \left[ \frac{1}{2} \partial^\mu h_{\alpha\beta}(\chi_0(\tau)) - \partial_\alpha h_\beta^\mu(\chi_0(\tau)) \right] \eqend{,}
\end{equation}
which can be integrated with the above boundary conditions to obtain
\begin{splitequation}
\label{calculation_prelim_firstordergeodesic}
\chi_1^\mu(\tau) &= - \frac{1}{2} h^\mu_\nu(x) v^\nu \tau + \int_0^\tau \int_0^{\tau'} v^\alpha v^\beta \left[ \frac{1}{2} \partial^\mu h_{\alpha\beta}(\tau'') - \partial_\alpha h_\beta^\mu(\tau'') \right] \total \tau'' \total \tau' \\
&= - \frac{1}{2} h^\mu_\nu(x) v^\nu \tau + \int_0^\tau (\tau - \tau') v^\alpha v^\beta \left[ \frac{1}{2} \partial^\mu h_{\alpha\beta}(\tau') - \partial_\alpha h_\beta^\mu(\tau') \right] \total \tau' \eqend{,}
\end{splitequation}
where we used the Cauchy formula for repeated integration~\cite{dlmf} in the second step, and write
\begin{equation}
h_{\alpha\beta}(\tau') \equiv h_{\alpha\beta}(\chi_0(\tau')) = h_{\alpha\beta}(x+v\tau') \eqend{.}
\end{equation}
It thus follows that the field entering the correlation function is, to first order in $\kappa$, given by
\begin{equation}
\phi(\chi(1)) = \phi(y) + \kappa \chi_1^\mu(1) \partial_\mu \phi(y) \eqend{,}
\end{equation}
where we have set $y^\mu \equiv x^\mu + v^\mu$ to simplify the expressions here and in the following. The second-order geodesic corrections follow the same calculational pattern and are presented in~\ref{appendix_geodesic}.

\begin{figure}[ht]
\includegraphics[width=0.22\textwidth]{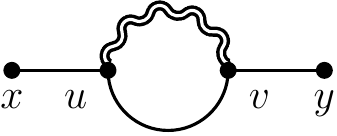}\hfil
\includegraphics[width=0.22\textwidth]{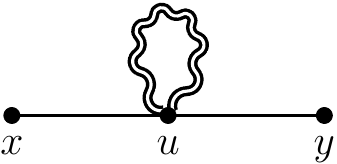}\hfil
\includegraphics[width=0.22\textwidth]{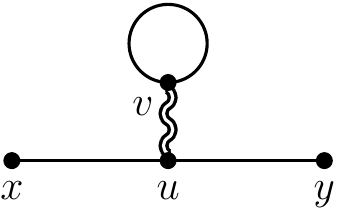}\hfil
\includegraphics[width=0.22\textwidth]{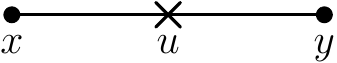}
\caption{Field-theoretic corrections to the scalar two-point function at fixed geodesic distance at order $\kappa^2$. Wiggly lines are gravitons, plain lines are scalars.}
\label{fig_feynman_1}
\end{figure}

\begin{figure}[ht]
\hfil\includegraphics[width=0.22\textwidth]{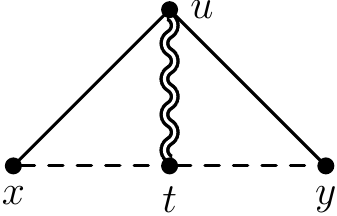}\hfil
\caption{Mixed geodesic/field-theoretic correction to the scalar two-point function at fixed geodesic distance at order $\kappa^2$. Wiggly lines are gravitons, plain lines are scalars. The point $t = x + (y-x) \tau$ is integrated over the geodesic, represented by the dashed line.}
\label{fig_feynman_2}
\end{figure}

\begin{figure}[ht]
\hfil\includegraphics[width=0.22\textwidth]{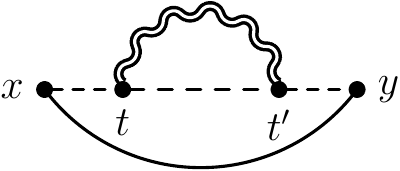}\hfil
\caption{Pure geodesic correction to the scalar two-point function at fixed geodesic distance at order $\kappa^2$. Wiggly lines are gravitons, plain lines are scalars. The points $t = x + (y-x) \tau$ and $t' = x + (y-x) \tau'$ are integrated over the geodesic, represented by the dashed line.}
\label{fig_feynman_3}
\end{figure}

To order $\kappa^2$, we have thus three contributions to the scalar two-point function at fixed geodesic distance: first, the usual field-theoretic corrections which are displayed in figure~\ref{fig_feynman_1}; second, terms where the end-point is corrected to first order and this is correlated with a first-order field-theoretic correction as shown in figure~\ref{fig_feynman_2}; and third, second-order corrections to the endpoint which are shown in figure~\ref{fig_feynman_3}. We will treat all those in separate subsections. Since all of these corrections involve exactly one graviton propagator, we can further organise the calculation by treating the tensor structures~\eqref{calculation_prelim_gravprop_tensors} one by one. To shorten the presentation, we will moreover only show explicit steps for $T^{(1)}$, and then just quote the results for the other tensor structures.


\subsection{Field-theoretic corrections: scalar self-energy}
\label{sec_calculation_selfenergy}

As explained, massless tadpoles vanish in dimensional regularisation, such that the second and third diagrams of figure~\ref{fig_feynman_1} do not contribute. The fourth diagram is the counterterm coming from $S_\text{CT}$~\eqref{calculation_prelim_action_ct}, whose contribution to the scalar two-point function reads
\begin{splitequation}
\label{calculation_selfenergy_deltact}
\mathcal{G}^\text{FT,CT}_0(x,y) &= \delta_{(\partial^2 \phi)^2} \int \expect{ \phi(x) \left( \partial^2 \phi(u) \right)^2 \phi(y) } \total^n u \\
&= - 2 \delta_{(\partial^2 \phi)^2} \int \partial_u^2 G_0(x,u) \partial_u^2 G_0(y,u) \total^n u \\
&= - 2 \delta_{(\partial^2 \phi)^2} \delta^n(x-y) \eqend{,}
\end{splitequation}
where we used the equation of motion~\eqref{calculation_prelim_eom_scalar}, and the fact that $\delta_{(\partial^2 \phi)^2} = \bigo{\kappa^2}$ (since no counterterm is necessary at tree level).

The contribution of the first diagram reads
\begin{splitequation}
\mathcal{G}^\text{FT,33}_0(x,y) &= \frac{\mathi}{2} \kappa^2 \left[ \expect{ \phi(x) \phi(y) S_1^{h\phi\phi} S_1^{h\phi\phi} } - \expect{ \phi(x) \phi(y) } \expect{ S_1^{h\phi\phi} S_1^{h\phi\phi} } \right] \\
&= \mathi \kappa^2 \iint G_0(x,u) G_0(y,v) \bigg[ \tau_{\mu\nu\alpha\beta} \tau_{\rho\sigma\gamma\delta} \partial^\alpha \partial^\gamma \left( \partial^\beta \partial^\delta G_0(u,v) G^{\mu\nu\rho\sigma}(u,v) \right) \\
&\hspace{6em}+ 2 \xi \tau_{\rho\sigma\gamma\delta} \partial^\gamma \left( \partial^\delta G_0(u,v) S_{\mu\nu} G^{\mu\nu\rho\sigma}(u,v) \right) \\
&\hspace{6em}+ \xi^2 G_0(u,v) S_{\mu\nu} S_{\rho\sigma} G^{\mu\nu\rho\sigma}(u,v) \bigg] \total^n u \total^n v \eqend{,}
\end{splitequation}
where we have integrated by parts all derivatives acting on the external propagators $G_0(x,u)$ and $G_0(y,v)$, and used that in addition to the obvious symmetries, the graviton propagator is symmetric under the interchange of the index pairs $(\mu\nu) \leftrightarrow (\rho\sigma)$ and only depends on $(u-v)^2$. The term in brackets is usually referred to as the scalar self-energy or self-mass, but for us it is of no use to treat it separately, i.e., without the external propagators. Performing a Fourier transform and using equation~\eqref{calculation_prelim_gravprop}, we obtain
\begin{splitequation}
\tilde{\mathcal{G}}^\text{FT,33}_0(p) &= \mathi \kappa^2 \sum_{i=1}^5 g_i \frac{1}{[ p^2 - \mathi 0 ]^2} \int \frac{1}{k^2 - \mathi 0} \frac{1}{(p-k)^2 - \mathi 0} \bigg[ \tau^{\mu\nu\alpha\beta} \tau^{\rho\sigma\gamma\delta} p_\alpha (p-k)_\beta p_\gamma (p-k)_\delta \\
&\quad+ \xi \left[ 2 \tau^{\rho\sigma\gamma\delta} p_\gamma (p-k)_\delta + \xi \left( k^\rho k^\sigma - \eta^{\rho\sigma} k^2 \right) \right] \left( k^\mu k^\nu - \eta^{\mu\nu} k^2 \right) \bigg] T^{(i)}_{\mu\nu\rho\sigma}(k) \frac{\total^n k}{(2\pi)^n} \eqend{,}
\end{splitequation}
which can now be evaluated for the five tensor structures~\eqref{calculation_prelim_gravprop_tensors} one by one. For $T^{(1)}$, we obtain [using also the definition of $\tau_{\mu\nu\rho\sigma}$~\eqref{calculation_prelim_tau}]
\begin{splitequation}
\tilde{\mathcal{G}}^\text{FT,33,(1)}_0(p) &= \mathi \kappa^2 \frac{1}{[ p^2 - \mathi 0 ]^2} \int \bigg[ \frac{p^2}{k^2 - \mathi 0} + \frac{n-2}{2} [ p^2 - (pk) ]^2 \frac{1}{k^2 - \mathi 0} \frac{1}{(p-k)^2 - \mathi 0} \\
&\qquad+ \xi \frac{4 (pk)^2}{k^2 - \mathi 0} \frac{1}{(p-k)^2 - \mathi 0} - \xi \frac{4 (pk)}{(p-k)^2 - \mathi 0} + 2 (n-3) \xi \frac{p^2 - (pk)}{(p-k)^2 - \mathi 0} \\
&\qquad- 2 (n-1) \xi^2 \frac{k^2}{(p-k)^2 - \mathi 0} \bigg] \frac{\total^n k}{(2\pi)^n} \eqend{.}
\end{splitequation}
The integral
\begin{equation}
\int \frac{p^2}{k^2 - \mathi 0} \frac{\total^n k}{(2\pi)^n} = p^2 \int \frac{1}{k^2 - \mathi 0} \frac{\total^n k}{(2\pi)^n}
\end{equation}
is a scaleless integral, which vanishes in dimensional regularisation~\cite{leibbrandt1975}. Similarly, by shifting the integration variable $k \to k-p$, one sees that also $\int [ (p-k)^2 - \mathi 0 ]^{-1} \total^n k / (2\pi)^n$ vanishes. Furthermore,
\begin{equation}
\int \frac{(pk)}{k^2 - \mathi 0} \frac{\total^n k}{(2\pi)^n} = p^\mu \int \frac{k_\mu}{k^2 - \mathi 0} \frac{\total^n k}{(2\pi)^n}
\end{equation}
vanishes by rotational symmetry, and by using that $2 (pk) = p^2 + k^2 - (p-k)^2$ we can reduce also the remaining terms, leading to
\begin{equation}
\tilde{\mathcal{G}}^\text{FT,33,(1)}_0(p) = \mathi \kappa^2 \left( \frac{n-2}{8} + \xi \right) \int \frac{1}{k^2 - \mathi 0} \frac{1}{(p-k)^2 - \mathi 0} \frac{\total^n k}{(2\pi)^n} \eqend{.}
\end{equation}
This is a standard 1-loop tabulated Feynman integral~\cite{smirnov2004} which can be explicitly given as function of $p^2$. However, for us it is more useful to note that
\begin{equation}
\label{calculation_selfenergy_propsquared}
G_0(x,y) G_0(x,y) = \iint \frac{1}{k^2 - \mathi 0} \frac{1}{(p-k)^2 - \mathi 0} \frac{\total^n k}{(2\pi)^n} \mathe^{\mathi p (x-y)} \frac{\total^n p}{(2\pi)^n} \eqend{,}
\end{equation}
such that inverting the Fourier transform we have
\begin{equation}
\mathcal{G}^\text{FT,33,(1)}_0(x,y) = \mathi \kappa^2 \left( \frac{n-2}{8} + \xi \right) G_0(x,y) G_0(x,y) \eqend{.}
\end{equation}

Similarly, for the contributions from the other tensor structures we obtain
\begin{equations}
\mathcal{G}^\text{FT,33,(2)}_0(x,y) &= \mathi \kappa^2 \frac{(n-2)^2}{16} G_0(x,y) G_0(x,y) \eqend{,} \\
\mathcal{G}^\text{FT,33,(3)}_0(x,y) &= \mathi \kappa^2 \frac{n-1}{2} \left( \frac{n-2}{4} - \xi \right) G_0(x,y) G_0(x,y) \eqend{,} \\
\mathcal{G}^\text{FT,33,(4)}_0(x,y) &= 0 \eqend{,} \\
\mathcal{G}^\text{FT,33,(5)}_0(x,y) &= \mathi \kappa^2 \frac{(n-1) (n-2)}{32} G_0(x,y) G_0(x,y) \eqend{,}
\end{equations}
where we also had to use equation~\eqref{appendix_feynman_scalarpowers}. Summing up, we have
\begin{splitequation}
\label{calculation_selfenergy_regresult}
\mathcal{G}^\text{FT,33}_0(x,y) &= \sum_{i=1}^5 g_i \mathcal{G}^\text{FT,33,(i)}_0(x,y) \\
&= \mathi \kappa^2 \left[ \xi - (n-1) (2+\beta) \left( \frac{n-4}{16} + \frac{1}{n-2} \xi + \frac{n}{32} \beta + \frac{n-2}{32} (\alpha-1) (2-\beta) \right) \right] \\
&\qquad\times G_0(x,y) G_0(x,y) \eqend{.}
\end{splitequation}
Since from the explicit expression~\eqref{calculation_prelim_g0} for the propagator in coordinate space it follows that
\begin{equation}
G_0(x,y) G_0(x,y) = - \frac{c_n^2}{[(x-y)^2]^{n-2}} \eqend{,}
\end{equation}
we see that the product of two propagators is too singular at coincidence to be a proper distribution in $n = 4$ dimensions. To renormalise it, we first extract a d'Alembertian using equation~\eqref{appendix_feynman_x2_dalembert_p} and then add an ``intelligent zero'' to obtain
\begin{splitequation}
G_0(x,y) G_0(x,y) &= - \frac{c_n^2}{2 (n-3) (n-4)} \partial^2 \frac{1}{[(x-y)^2]^{n-3}} \\
&= - \frac{c_n^2}{2 (n-3) (n-4)} \partial^2 \left[ \frac{1}{[(x-y)^2]^{n-3}} - \frac{\mu^{n-4}}{[(x-y)^2]^\frac{n-2}{2}} \right] \\
&\qquad- \frac{\mathi c_n}{2 (n-3) (n-4)} \mu^{n-4} \partial^2 G_0(x,y) \eqend{,}
\end{splitequation}
where $\mu$ is an arbitrary scale (the renormalisation scale) introduced to make the equation dimensionally correct. Now the terms in the first line have a well-defined limit as $n \to 4$, while the second line reduces to a local term on account of the equation of motion~\eqref{calculation_prelim_eom_scalar}. Expanding the constants around $n = 4$ [using also equation~\eqref{calculation_prelim_cn_def}], it follows that
\begin{splitequation}
G_0(x,y) G_0(x,y) &= \frac{1}{64 \pi^4} \partial^2 \left[ \frac{\ln[ \mu^2 (x-y)^2 ]}{(x-y)^2} \right] \\
&\qquad- \frac{\mathi}{16 \pi^2} \left[ \frac{2}{n-4} - 2 + 2 \ln \mu - \gamma - \ln \pi + \bigo{n-4} \right] \delta^n(x-y) \eqend{.}
\end{splitequation}
Let us denote
\begin{equation}
\label{calculation_selfenergy_h0_def}
H^{(k)}_0(x;\mu) \equiv \frac{\mathi}{64 \pi^4} \partial^2 \left[ \frac{\ln^k( \mu^2 x^2 )}{x^2} \right] \eqend{,}
\end{equation}
which now is a well-defined distribution in four dimensions. Taking then the counterterm $\delta_{(\partial^2 \phi)^2}$~\eqref{calculation_selfenergy_deltact} to be
\begin{splitequation}
\label{calculation_selfenergy_ct_value}
\delta_{(\partial^2 \phi)^2} &= \frac{\kappa^2}{32 \pi^2} \bigg[ \left[ \xi - \frac{3}{16} (2+\beta) \left( 8 \xi + 2 \beta + (\alpha-1) (2-\beta) \right) \right] \left( \frac{2}{n-4} + 2 \ln \bar{\mu} \right) \\
&\qquad\qquad- \frac{2+\beta}{16} \left( - 8 \xi + 7 \beta + 5 (\alpha-1) (2-\beta) + 6 \right) \bigg] + \delta_{(\partial^2 \phi)^2}^\text{fin}(\mu)
\end{splitequation}
with an arbitrary finite constant $\delta_{(\partial^2 \phi)^2}^\text{fin}(\mu)$ and
\begin{equation}
\label{calculation_selfenergy_mubar}
\bar{\mu} \equiv \frac{\mu}{\sqrt{\pi \mathe^{\gamma+2}}} \eqend{,}
\end{equation}
the sum
\begin{splitequation}
\label{calculation_selfenergy_result}
\mathcal{G}^\text{FT,33}_0(x,y) + \mathcal{G}^\text{FT,CT}_0(x,y) &= \kappa^2 \left[ \xi - \frac{3}{16} (2+\beta) \left( 8 \xi + 2 \beta + (\alpha-1) (2-\beta) \right) \right] H^{(1)}_0(x-y;\mu) \\
&\quad- 2 \delta_{(\partial^2 \phi)^2}^\text{fin}(\mu) \delta^n(x-y)
\end{splitequation}
is finite in $n = 4$ dimensions. Since the bare action is independent of the regularisation scale $\mu$, for the $\mu$-dependence of $\delta_{(\partial^2 \phi)^2}^\text{fin}(\mu)$ we calculate from equation~\eqref{calculation_selfenergy_ct_value}
\begin{equation}
\mu \frac{\total}{\total \mu} \delta_{(\partial^2 \phi)^2}^\text{fin}(\mu) = - \frac{\kappa^2}{16 \pi^2} \left[ \xi - \frac{3}{16} (2+\beta) \left( 8 \xi + 2 \beta + (\alpha-1) (2-\beta) \right) \right] \eqend{.}
\end{equation}
Note that both the counterterm~\eqref{calculation_selfenergy_ct_value} and the final result~\eqref{calculation_selfenergy_result} depend on the gauge, and vanish in the analogue of Feynman gauge, $\beta = -2$, for minimal coupling of the scalar field to gravity, $\xi = 0$.

We can compare this result with the flat-space limit of de~Sitter calculations for the minimally and conformally coupled scalar~\cite{kahyawoodard2007,borankahyapark2014,borankahyapark2017}, which use a gauge that in the flat-space limit reduces to Feynman gauge $\alpha = 1$, $\beta = -2$. Both the sum~\eqref{calculation_selfenergy_result} and the higher-derivative counterterm~\eqref{calculation_selfenergy_ct_value} then vanish for minimal coupling $\xi = 0$, consistent with the result of ref.~\cite{kahyawoodard2007}. For conformal coupling $\xi = (n-2)/[4(n-1)] = 1/6$ in $n = 4$ dimensions, the sum~\eqref{calculation_selfenergy_result} reads in Feynman gauge (ignoring the finite part of the counterterm)
\begin{splitequation}
\mathi \frac{\kappa^2}{384 \pi^4} \partial^2 \left[ \frac{\ln( \mu^2 x^2 )}{x^2} \right] \eqend{,}
\end{splitequation}
and the higher-derivative counterterm~\eqref{calculation_selfenergy_ct_value} is
\begin{equation}
\delta_{(\partial^2 \phi)^2} = \frac{\kappa^2}{192 \pi^2} \frac{2}{n-4} + \bigo{(n-4)^0} \eqend{.}
\end{equation}
In contrast, the results of refs.~\cite{borankahyapark2014,borankahyapark2017} read in the flat-space limit
\begin{equation}
\mathi \frac{\kappa^2}{32 \pi^4} \left( \frac{1}{6} + 18 \right) \partial^2 \left[ \frac{\ln( \mu^2 x^2 )}{x^2} \right]
\end{equation}
after convoluting their self-energy with two external massless scalar propagators, and
\begin{equation}
\delta_{(\partial^2 \phi)^2} = \frac{\kappa^2}{2304 \pi^2} \frac{2}{n-4} + \bigo{(n-4)^0}
\end{equation}
after correcting for the different normalisation, which do \emph{not} agree. The cause of the discrepancy is not known at present.

\subsection{First-order geodesic corrections}
\label{sec_calculation_firstgeodesic}

The diagram of figure~\ref{fig_feynman_2} translates into the following contribution to the two-point function at fixed geodesic distance:
\begin{splitequation}
\mathcal{G}_0^{\text{FG}}(x,y) &= \kappa^2 \expect{ \phi(x) \chi_1^\mu(1) \partial_\mu \phi(y) S_1^{h\phi\phi} } \\
&= - \mathi \kappa^2 \int \left[ \tau_{\rho\sigma\gamma\delta} \partial^\gamma G_0(x,u) \partial_\mu \partial^\delta G_0(y,u) - \xi S^u_{\rho\sigma} \left( G_0(x,u) \partial_\mu G_0(y,u) \right) \right] \\
&\quad\times \bigg[ \int_0^1 (1-\tau) (x-y)_\alpha (x-y)_\beta \left[ \frac{1}{2} \partial^\mu G^{\alpha\beta\rho\sigma}(\tau,u) - \partial^\alpha G^{\mu\beta\rho\sigma}(\tau,u) \right] \total \tau \\
&\qquad\quad+ \frac{1}{2} (x-y)_\nu G^{\mu\nu\rho\sigma}(u,x) \bigg] \total^n u \eqend{.}
\end{splitequation}
Let us first treat the term without a $\tau$ integration. Performing a Fourier transform and using equation~\eqref{calculation_prelim_gravprop}, we obtain
\begin{splitequation}
\mathcal{G}_0^{\text{FG},0}(x,y) &= - \frac{1}{2} \kappa^2 \sum_{i=1}^5 g_i (x-y)^\nu \int \frac{1}{p^2 - \mathi 0} \int \frac{1}{k^2 - \mathi 0} \frac{1}{(p-k)^2 - \mathi 0} \\
&\quad\times \left[ \tau^{\rho\sigma\gamma\delta} (k-p)_\gamma p^\mu p_\delta - \xi \left( k^\rho k^\sigma - \eta^{\rho\sigma} k^2 \right) p^\mu \right] T^{(i)}_{\mu\nu\rho\sigma}(k) \frac{\total^n k}{(2\pi)^n} \mathe^{\mathi p (x-y)} \frac{\total^n p}{(2\pi)^n} \eqend{,}
\end{splitequation}
which can now be evaluated for the five tensor structures~\eqref{calculation_prelim_gravprop_tensors} one by one. The calculation is mostly analogous to the purely field-theoretic corrections, with only two exceptions: first, there is a tensor factor $k^\nu$ which can be converted into a factor $p^\nu$ using equation~\eqref{appendix_feynman_tensorfactors} and then into a derivative with respect to $x^\nu$; second, the final result is thus not just proportional to $G_0(x,y) G_0(x,y)$, but to
\begin{equation}
\label{calculation_firstgeodesic_eulerprop}
(x-y)_\nu \partial^\nu \left[ G_0(x,y) G_0(x,y) \right] = - 2 (n-2) G_0(x,y) G_0(x,y)
\end{equation}
because of the explicit form of $G_0$~\eqref{calculation_prelim_g0}. Since these are only minor modifications, we simply state the final results:
\begin{equations}
\mathcal{G}_0^{\text{FG},0,(1)}(x,y) &= \mathi \kappa^2 \frac{n-2}{2} (1+\xi) G_0(x,y) G_0(x,y) \eqend{,} \\
\mathcal{G}_0^{\text{FG},0,(2)}(x,y) &= - \mathi \kappa^2 \frac{(n-2)^2}{4} G_0(x,y) G_0(x,y) \eqend{,} \\
\mathcal{G}_0^{\text{FG},0,(3)}(x,y) &= - \mathi \kappa^2 \frac{(n-2)}{16} \left[ (8-n) (n-2) + 4 (n-1) \xi \right] G_0(x,y) G_0(x,y) \eqend{,} \\
\mathcal{G}_0^{\text{FG},0,(4)}(x,y) &= - \mathi \kappa^2 \frac{(n-2) (n-3)}{2} G_0(x,y) G_0(x,y) \eqend{,} \\
\mathcal{G}_0^{\text{FG},0,(5)}(x,y) &= \mathi \kappa^2 \frac{(n-2) (n^2-9n+16)}{32} G_0(x,y) G_0(x,y) \eqend{.}
\end{equations}

The terms with a $\tau$ integration are much harder. Passing to Fourier space and using equation~\eqref{calculation_prelim_gravprop}, we obtain
\begin{splitequation}
\mathcal{G}_0^{\text{FG},\tau}(x,y) &= - \mathi \kappa^2 \sum_{i=1}^5 g_i \int \frac{1}{p^2 - \mathi 0} \int_0^1 (1-\tau) \int \frac{1}{k^2 - \mathi 0} \frac{1}{(p-k)^2 - \mathi 0} \\
&\qquad\times \left[ \tau^{\rho\sigma\gamma\delta} (k-p)_\gamma p_\delta - \xi \left( k_\rho k_\sigma - \eta_{\rho\sigma} k^2 \right) \right] (x-y)^\alpha (x-y)^\beta \\
&\qquad\times \left[ \frac{1}{2} (pk) T^{(i)}_{\alpha\beta\rho\sigma}(k) - k_\alpha p^\mu T^{(i)}_{\mu\beta\rho\sigma}(k) \right] \mathe^{- \mathi k (x-y) \tau} \frac{\total^n k}{(2\pi)^n} \total \tau \mathe^{\mathi p (x-y)} \frac{\total^n p}{(2\pi)^n} \eqend{.}
\end{splitequation}
Note that the $k$ and $\tau$ integrations are entangled, in the sense that there is a factor $\exp\left[ - \mathi k (x-y) \tau \right]$ which does not permit us to perform the $k$ integral as for the field-theoretic contributions, and which gives unpleasant inverse powers of $k(x-y)$ if we perform the $\tau$ integration first. Similar integrals appear in the perturbative calculation of Wilson loops and invariant quark correlation functions~\cite{stefanis_thesis,stefanis1984}, and to untangle, we perform the shift $p \to p + \tau k$ which results in
\begin{splitequation}
\mathcal{G}_0^{\text{FG},\tau}(x,y) &= - \mathi \kappa^2 \sum_{i=1}^5 g_i \int \int_0^1 (1-\tau) \int \frac{1}{k^2 - \mathi 0} \frac{1}{(p + \tau k)^2 - \mathi 0} \frac{1}{(p - (1-\tau) k)^2 - \mathi 0} \\
&\qquad\times \left[ \tau^{\rho\sigma\gamma\delta} \left( (1-\tau) k_\gamma - p_\gamma \right) \left( p_\delta + \tau k_\delta \right) - \xi \left( k^\rho k^\sigma - \eta^{\rho\sigma} k^2 \right) \right] (x-y)^\alpha (x-y)^\beta \\
&\qquad\times \left[ \frac{1}{2} [ (pk) + \tau k^2 ] T^{(i)}_{\alpha\beta\rho\sigma}(k) - k_\alpha \left( p^\mu + \tau k^\mu \right) T^{(i)}_{\mu\beta\rho\sigma}(k) \right] \frac{\total^n k}{(2\pi)^n} \total \tau \mathe^{\mathi p (x-y)} \frac{\total^n p}{(2\pi)^n} \eqend{.}
\end{splitequation}
The $k$ integral can now be performed, and afterwards we can integrate over $\tau$. Again, this can be done separately for the five tensor structures~\eqref{calculation_prelim_gravprop_tensors}, and we only show the explicit steps for $T^{(1)}$. Using the definition of $\tau_{\mu\nu\rho\sigma}$~\eqref{calculation_prelim_tau}, we get
\begin{splitequation}
\mathcal{G}_0^{\text{FG},\tau,(1)}(x,y) &= \mathi \kappa^2 \int \int_0^1 (1-\tau) \int \frac{1}{k^2 - \mathi 0} \frac{1}{(p + \tau k)^2 - \mathi 0} \frac{1}{(p - (1-\tau) k)^2 - \mathi 0} \\
&\qquad\times \bigg[ (1-\tau) \left( p^2 + \tau (pk) \right) k_\alpha k_\beta - \left( p^2 + (pk) + (1-\tau) \tau k^2 \right) k_\alpha p_\beta \\
&\qquad\quad+ \frac{1}{2} \left( (1-\tau) \tau k^2 + (1-2\tau) (pk) - p^2 \right) \left( (pk) + \tau k^2 \right) \eta_{\alpha\beta} \\
&\qquad\quad+ \left( (pk) + \tau k^2 \right) p_\alpha p_\beta + \xi \left( \tau k^2 - (pk) \right) k_\alpha k_\beta + 2 \xi k^2 k_\alpha p_\beta \\
&\qquad\quad- \xi [ (pk) + \tau k^2 ] k^2 \eta_{\alpha\beta} \bigg] (x-y)^\alpha (x-y)^\beta \frac{\total^n k}{(2\pi)^n} \total \tau \mathe^{\mathi p (x-y)} \frac{\total^n p}{(2\pi)^n} \eqend{.}
\end{splitequation}

To be able to calculate the $k$ integral, we need to reduce the product of three propagators in the first line to two. This is immediate for the terms containing a $k^2$, while for the other terms we use
\begin{equations}
p^2 &= (1-\tau) (p + \tau k)^2 + \tau (p - (1-\tau) k)^2 - \tau (1-\tau) k^2 \eqend{,} \\
(pk) &= \frac{1}{2} \left[ (p + \tau k)^2 - (p - (1-\tau) k)^2 + (1-2\tau) k^2 \right] \eqend{.}
\end{equations}
We also take advantage of the fact that the region of integration is invariant under the exchange $\tau \leftrightarrow (1-\tau)$ to simplify the integrand, and obtain
\begin{splitequation}
\mathcal{G}_0^{\text{FG},\tau,(1)}(x,y) &= \frac{1}{8} \mathi \kappa^2 \int \int_0^1 \int \bigg[ 4 \bigg[ \tau \left( 1 - \tau + 2 \tau^2 \right) k_\alpha k_\beta + \left( 1 - 2 \tau + 4 \tau^2 \right) k_\alpha p_\beta \\
&\qquad\qquad\quad- (1-2\tau) \left( p_\alpha p_\beta + \frac{1}{2} (pk) \eta_{\alpha\beta} - \xi k_\alpha k_\beta \right) \bigg] \frac{1}{k^2 - \mathi 0} \frac{1}{(p + \tau k)^2 - \mathi 0} \\
&\quad+ \bigg[ - 4 \tau \left( (1-\tau) \tau - (3-4\tau) \xi \right) k_\alpha k_\beta - 4 \tau (1-2\tau+4\xi) k_\alpha p_\beta + 4 \tau p_\alpha p_\beta \\
&\qquad\qquad\quad+ (1-4\xi) \tau k^2 \eta_{\alpha\beta} \bigg] \frac{1}{(p + \tau k)^2 - \mathi 0} \frac{1}{(p - (1-\tau) k)^2 - \mathi 0} \bigg] \\
&\quad\times (x-y)^\alpha (x-y)^\beta \frac{\total^n k}{(2\pi)^n} \total \tau \mathe^{\mathi p (x-y)} \frac{\total^n p}{(2\pi)^n} \eqend{.}
\end{splitequation}
The tensor factors $k^\mu$ can now be reduced with the help of equations~\eqref{appendix_feynman_tensorfactors2} and~\eqref{appendix_feynman_tensorfactors3}, which (again using that the region of integration is invariant under the exchange $\tau \leftrightarrow (1-\tau)$ to simplify the integrand) leads to
\begin{splitequation}
\mathcal{G}_0^{\text{FG},\tau,(1)}(x,y) &= \frac{1}{8} \mathi \kappa^2 \int \int_0^1 \int \bigg[ - \frac{(n-2) + n \tau (1-2\tau)}{(n-1) \tau} \frac{p_\alpha p_\beta}{k^2 - \mathi 0} \frac{1}{(p + \tau k)^2 - \mathi 0} \\
&\quad+ \frac{(n-2) - (2n-3) \tau - 2 \tau^2}{(n-1) \tau} \eta_{\alpha\beta} p^2 \frac{1}{k^2 - \mathi 0} \frac{1}{(p + \tau k)^2 - \mathi 0} \\
&\quad+ \frac{(1-2\tau) \xi}{(n-1) \tau^2} \left( n p_\alpha p_\beta - \eta_{\alpha\beta} p^2 \right) \frac{1}{k^2 - \mathi 0} \frac{1}{(p + \tau k)^2 - \mathi 0} \\
&\quad+ \frac{n-2}{(n-1) \tau} \left( p_\alpha p_\beta - \eta_{\alpha\beta} p^2 \right) \frac{1}{(p + \tau k)^2 - \mathi 0} \frac{1}{(p - (1-\tau) k)^2 - \mathi 0} \\
&\quad- \frac{n - 4 (n+1) \tau (1-\tau)}{2 (n-1) \tau^2 (1-\tau)^2} \xi \frac{p_\alpha p_\beta}{(p + \tau k)^2 - \mathi 0} \frac{1}{(p - (1-\tau) k)^2 - \mathi 0} \\
&\quad+ \frac{1 + 4 (n-3) \tau (1-\tau)}{2 (n-1) \tau^2 (1-\tau)^2} \xi \frac{\eta_{\alpha\beta} p^2}{(p + \tau k)^2 - \mathi 0} \frac{1}{(p - (1-\tau) k)^2 - \mathi 0} \\
&\quad\bigg] (x-y)^\alpha (x-y)^\beta \frac{\total^n k}{(2\pi)^n} \total \tau \mathe^{\mathi p (x-y)} \frac{\total^n p}{(2\pi)^n} \eqend{.}
\end{splitequation}
Converting $p^\alpha$ into a derivative with respect to $x^\alpha$ and taking it out of the integral, the $k$ integral can be done using equations~\eqref{appendix_feynman_geodesic_integrals} and~\eqref{calculation_selfenergy_propsquared}, which gives
\begin{splitequation}
\mathcal{G}_0^{\text{FG},\tau,(1)}(x,y) &= \frac{1}{8 (n-1)} \mathi \kappa^2 \int_0^1 \bigg[ - \frac{(n-2) - (2n-3) \tau - 2 \tau^2}{\tau^{n-1}} (x-y)^2 \partial^2 \left[ G_0(x,y) G_0(x,y) \right] \\
&\qquad+ \frac{(n-2) + n \tau (1-2\tau)}{\tau^{n-1}} (x-y)^\alpha (x-y)^\beta \partial_\alpha \partial_\beta \left[ G_0(x,y) G_0(x,y) \right] \\
&\qquad- \frac{n-2}{\tau^{n-1} (1-\tau)^{n-2}} \left( (x-y)^\alpha (x-y)^\beta \partial_\alpha \partial_\beta - (x-y)^2 \partial^2 \right) \left[ G_0(x,y) G_0(x,y) \right] \\
&\qquad- \frac{1-2\tau}{\tau^n} \xi \left( n (x-y)^\alpha (x-y)^\beta \partial_\alpha \partial_\beta - (x-y)^2 \partial^2 \right) \left[ G_0(x,y) G_0(x,y) \right] \\
&\qquad+ \frac{n - 4 (n+1) \tau (1-\tau)}{2 \tau^n (1-\tau)^n} \xi (x-y)^\alpha (x-y)^\beta \partial_\alpha \partial_\beta \left[ G_0(x,y) G_0(x,y) \right] \\
&\qquad- \frac{1 + 4 (n-3) \tau (1-\tau)}{2 \tau^n (1-\tau)^n} \xi (x-y)^2 \partial^2 \left[ G_0(x,y) G_0(x,y) \right] \bigg] \total \tau \eqend{.}
\end{splitequation}
From equations~\eqref{calculation_prelim_g0} and~\eqref{calculation_firstgeodesic_eulerprop} we obtain
\begin{equations}
(x-y)^\alpha (x-y)^\beta \partial_\alpha \partial_\beta \left[ G_0(x,y) G_0(x,y) \right] &= 2 (n-2) (2n-3) G_0(x,y) G_0(x,y) \eqend{,} \\
(x-y)^2 \partial^2 \left[ G_0(x,y) G_0(x,y) \right] &= 2 (n-2)^2 G_0(x,y) G_0(x,y) \eqend{,}
\end{equations}
and the $\tau$ integral can be done using
\begin{equation}
\int_0^1 \tau^\alpha (1-\tau)^\beta \total \tau = \frac{\Gamma(\alpha+1) \Gamma(\beta+1)}{\Gamma(\alpha+\beta+2)} \eqend{.}
\end{equation}
It then follows that
\begin{splitequation}
\mathcal{G}_0^{\text{FG},\tau,(1)}(x,y) &= \mathi \kappa^2 \bigg[ \frac{(n-2)}{4} \left( \frac{\Gamma(3-n) \Gamma(3-n)}{\Gamma(5-2n)} - \frac{(n^2-13n+24)}{(n-3) (n-4)} \right) \\
&\qquad\qquad+ \frac{n}{2} \xi \left( \frac{\Gamma(2-n) \Gamma(3-n)}{\Gamma(4-2n)} - 1 \right) \bigg] G_0(x,y) G_0(x,y) \eqend{.}
\end{splitequation}
Note that this is divergent for $n = 4$, even for separated points $x \neq y$ because already the prefactor diverges. Concretely, expanding around $n = 4$ we obtain
\begin{splitequation}
\mathcal{G}_0^{\text{FG},\tau,(1)}(x,y) &= \mathi \kappa^2 \bigg[ \frac{12 (1+4\xi)}{n-4} + \frac{5}{2} \left( 5 + 36 \xi \right) + \frac{1}{4} \left( 17 - 4 \pi^2 + 14 \xi - 32 \pi^2 \xi \right) (n-4) \\
&\qquad\qquad+ \bigo{(n-4)^2} \bigg] G_0(x,y) G_0(x,y) \eqend{.}
\end{splitequation}
In the same way, the result for the other tensor structures is given by
\begin{splitequation}
\mathcal{G}_0^{\text{FG},\tau,(2)}(x,y) &= \mathi \kappa^2 \frac{n-2}{8} \bigg[ \frac{(n-1) (3n-8)}{n-3} \\
&\qquad\qquad+ n [ (n-2) - 4 (n-1) \xi ] \frac{\Gamma(2-n) \Gamma(3-n)}{\Gamma(5-2n)} \bigg] G_0(x,y) G_0(x,y) \\
&= - \mathi \kappa^2 \bigg[ \frac{12 (1-6\xi)}{n-4} + 2 (13-81\xi) + \frac{1}{4} \left( 43 - 8 \pi^2 - 208 \xi + 48 \pi^2 \xi \right) (n-4) \\
&\qquad\qquad+ \bigo{(n-4)^2} \bigg] G_0(x,y) G_0(x,y) \eqend{,}
\end{splitequation}
\begin{equation}
\mathcal{G}_0^{\text{FG},\tau,(3)}(x,y) = \mathi \kappa^2 \bigg[ - \frac{(n-2)^2 (n^2-11n+16)}{16 (n-3)} + \frac{n (n-1)}{4} \xi \bigg] G_0(x,y) G_0(x,y) \eqend{,}
\end{equation}
\begin{equation}
\mathcal{G}_0^{\text{FG},\tau,(4)}(x,y) = \mathi \kappa^2 \frac{(n-2)^2}{2} G_0(x,y) G_0(x,y) \eqend{,}
\end{equation}
\begin{equation}
\mathcal{G}_0^{\text{FG},\tau,(5)}(x,y) = \mathi \kappa^2 \frac{(n-2)^2 (5-n)}{32} G_0(x,y) G_0(x,y) \eqend{.}
\end{equation}

\subsection{Second-order geodesic corrections}
\label{sec_calculation_secondgeodesic}

The diagram of figure~\ref{fig_feynman_3} translates into the following contribution to the two-point function at fixed geodesic distance:
\begin{splitequation}
\mathcal{G}_0^{\text{SG}}(x,y) &= - \mathi \kappa^2 \expect{ \phi(x) \chi_2^\mu(1) \partial_\mu \phi(y) } - \frac{\mathi}{2} \kappa^2 \expect{ \phi(x) \chi_1^\mu(1) \chi_1^\nu(1) \partial_\mu \partial_\nu \phi(y) } \\
&= \kappa^2 \partial^y_\mu G_0(x,y) \expect{ \chi_2^\mu(1) } + \frac{1}{2} \kappa^2 \partial^y_\mu \partial^y_\nu G_0(x,y) \expect{ \chi_1^\mu(1) \chi_1^\nu(1) } \eqend{,}
\end{splitequation}
where all the graviton operators are contained in $\chi_1^\mu$ and $\chi_2^\mu$, whose explicit expressions are given in~\ref{appendix_geodesic}. Since massless tadpoles vanish in dimensional regularisation, we can drop all terms where two graviton operators are taken at the same point. Furthermore, the usual time-ordering translates into a path ordering $\mathcal{P}$ for the gravitons along the geodesic, which needs to be taken into account in the expectation value $\expect{ \chi_1^\mu(1) \chi_1^\nu(1) }$ (the other correction is already ordered).

Let us start with the $\chi_2^\mu$ correction, passing to Fourier space and using equation~\eqref{calculation_prelim_gravprop} to obtain
\begin{splitequation}
\expect{ \dot \chi_2^\mu(\tau) } &= - \frac{\mathi}{2} (x-y)^\beta (x-y)^\gamma (x-y)^\delta \sum_{i=1}^5 g_i \int_0^\tau \int_0^{\tau'} \int \frac{1}{k^2 - \mathi 0} \left( k^\alpha \delta_\gamma^\rho - 2 k_\gamma \eta^{\alpha\rho} \right) \eta^{\mu\sigma} \\
&\qquad\quad\times \left[ k_\alpha T^{(i)}_{\sigma\beta\rho\delta}(k) + k_\beta T^{(i)}_{\sigma\alpha\rho\delta}(k) - k_\sigma T^{(i)}_{\alpha\beta\rho\delta}(k) \right] \mathe^{- \mathi k (x-y) (\tau'-\tau'')} \frac{\total^n k}{(2\pi)^n} \total \tau'' \total \tau' \\
&\quad+ \frac{1}{4} (x-y)^\alpha (x-y)^\beta (x-y)^\gamma (x-y)^\delta \sum_{i=1}^5 g_i \int_0^\tau \int_0^{\tau'} (\tau' - \tau'') \int \frac{1}{k^2 - \mathi 0} \\
&\qquad\quad\times \left( k^2 \delta^\rho_\gamma - 2 k_\gamma k^\rho \right) \left( k^\mu \delta^\sigma_\alpha - 2 k_\alpha \eta^{\sigma\mu} \right) T^{(i)}_{\sigma\beta\rho\delta}(k) \mathe^{- \mathi k (x-y) (\tau'-\tau'')} \frac{\total^n k}{(2\pi)^n} \total \tau'' \total \tau' \eqend{.}
\end{splitequation}
To disentangle the $\tau$ and $k$ integrals, we rescale $k \to k/(\tau'-\tau'')$. Since all of the five tensor structures~\eqref{calculation_prelim_gravprop_tensors} are homogeneous of degree zero in $k$, they do not change under the rescaling, and the $\tau$ integrals can be performed easily. Again, we can treat each of the tensor structures separately, and obtain
\begin{splitequation}
\mathcal{G}_0^{\text{SG},2,(i)}(x,y) &= - \kappa^2 \partial_\mu G_0(x,y) \int_0^\tau \expect{ \dot \chi_2^\mu(\tau) }^{(i)} \total \tau \\
&= \frac{\kappa^2}{4 (n-1) (n-2) (n-3)} \partial_\mu G_0(x,y) \bigg[ - 2 \mathi (x-y)^\beta (x-y)^\gamma (x-y)^\delta \\
&\qquad\quad\times \int \frac{1}{k^2 - \mathi 0} \left( k^\alpha \delta_\gamma^\rho - 2 k_\gamma \eta^{\alpha\rho} \right) \eta^{\mu\sigma} \\
&\qquad\quad\times \left[ k_\alpha T^{(i)}_{\sigma\beta\rho\delta}(k) + k_\beta T^{(i)}_{\sigma\alpha\rho\delta}(k) - k_\sigma T^{(i)}_{\alpha\beta\rho\delta}(k) \right] \mathe^{- \mathi k (x-y)} \frac{\total^n k}{(2\pi)^n} \\
&\quad+ (x-y)^\alpha (x-y)^\beta (x-y)^\gamma (x-y)^\delta \int \frac{1}{k^2 - \mathi 0} \\
&\qquad\quad\times \left( k^2 \delta^\rho_\gamma - 2 k_\gamma k^\rho \right) \left( k^\mu \delta^\sigma_\alpha - 2 k_\alpha \eta^{\sigma\mu} \right) T^{(i)}_{\sigma\beta\rho\delta}(k) \mathe^{- \mathi k (x-y)} \frac{\total^n k}{(2\pi)^n} \bigg] \eqend{.}
\end{splitequation}
For the first tensor structure~\eqref{calculation_prelim_gravprop_tensors}, performing the tensor algebra it follows that
\begin{splitequation}
\mathcal{G}_0^{\text{SG},2,(1)}(x,y) &= - \kappa^2 \partial_\mu G_0(x,y) \int_0^1 \expect{ \dot \chi_2^\mu(\tau) }^{(i)} \total \tau \\
&= \frac{\kappa^2}{(n-1) (n-2) (n-3)} (x-y)^\mu \partial_\mu G_0(x,y) \int \frac{1}{k^2 - \mathi 0} \bigg[ - \mathi k^2 (x-y)^2 \\
&\qquad\quad- k^2 [ k (x-y) ] (x-y)^2 + \mathi (n+1) [ k (x-y) ]^2 + [ k (x-y) ]^3 \bigg] \mathe^{- \mathi k (x-y)} \frac{\total^n k}{(2\pi)^n} \\
&\quad+ \frac{\kappa^2}{2 (n-1) (n-2) (n-3)} (x-y)^2 \partial_\mu G_0(x,y) \\
&\qquad\quad\times \int \frac{k^\mu}{k^2 - \mathi 0} \bigg[ k^2 (x-y)^2 - 2 \mathi (n-1) [ k (x-y) ] \bigg] \mathe^{- \mathi k (x-y)} \frac{\total^n k}{(2\pi)^n} \eqend{.}
\end{splitequation}
We then convert all $k^\alpha$ into derivatives, which can be taken out of the integral. The remaining integral is just the massless propagator, and we obtain [using also the relation~\eqref{calculation_prelim_eom_scalar}]
\begin{splitequation}
\mathcal{G}_0^{\text{SG},2,(1)}(x,y) &= - \frac{\mathi \kappa^2}{(n-1) (n-2) (n-3)} (x-y)^2 (x-y)^\mu \partial_\mu G_0(x,y) \left[ 1 + (x-y)^\alpha \partial_\alpha \right] \delta^n(x-y) \\
&\quad+ \frac{\mathi \kappa^2}{(n-1) (n-2) (n-3)} (x-y)^\mu \partial_\mu G_0(x,y) \bigg[ (n+1) (x-y)^\alpha (x-y)^\beta \partial_\alpha \partial_\beta + (x-y)^\alpha (x-y)^\beta (x-y)^\gamma \partial_\alpha \partial_\beta \partial_\gamma \bigg] G_0(x,y) \\
&\quad+ \frac{\mathi \kappa^2}{2 (n-1) (n-2) (n-3)} (x-y)^2 \partial_\mu G_0(x,y) (x-y)^2 \partial^\mu \delta^n(x-y) \\
&\quad- \frac{\mathi \kappa^2}{(n-2) (n-3)} (x-y)^2 \partial_\mu G_0(x,y) (x-y)^\alpha \partial_\alpha \partial^\mu G_0(x,y) \eqend{.}
\end{splitequation}
With the explicit form of $G_0$~\eqref{calculation_prelim_g0}, we also obtain
\begin{equations}
(x-y)^\alpha \partial_\alpha G_0(x,y) &= - (n-2) G_0(x,y) \eqend{,} \\
(x-y)^2 \partial_\mu G_0(x,y) \partial^\mu G_0(x,y) &= (n-2)^2 G_0(x,y) G_0(x,y) \eqend{,}
\end{equations}
and using that $n$-dependent powers of $(x-y)$ vanish at coincidence in dimensional regularisation, after some algebra it follows that
\begin{equation}
\mathcal{G}_0^{\text{SG},2,(1)}(x,y) = \mathi \kappa^2 \frac{(n-2)^2}{n-3} G_0(x,y) G_0(x,y) \eqend{.}
\end{equation}
The result for the other tensor structures can be calculated in the same way, using in addition that
\begin{equation}
\int \frac{1}{[ k^2 - \mathi 0 ]^2} \mathe^{- \mathi k (x-y)} \frac{\total^n k}{(2\pi)^n} = \partial^{-2} G_0(x,y) = - \frac{1}{2 (n-4)} (x-y)^2 G_0(x,y) \eqend{,}
\end{equation}
and we obtain
\begin{equations}
\mathcal{G}_0^{\text{SG},2,(2)}(x,y) &= \mathi \kappa^2 \frac{(n-2)^2}{2 (n-3)} G_0(x,y) G_0(x,y) \eqend{,} \\
\mathcal{G}_0^{\text{SG},2,(3)}(x,y) &= - \mathi \kappa^2 \frac{(n-2)^2}{4} G_0(x,y) G_0(x,y) \eqend{,} \\
\mathcal{G}_0^{\text{SG},2,(4)}(x,y) &= - \mathi \kappa^2 \frac{(n-2)^2}{2} G_0(x,y) G_0(x,y) \eqend{,} \\
\mathcal{G}_0^{\text{SG},2,(5)}(x,y) &= - \mathi \kappa^2 \frac{(n-2)^2}{8} G_0(x,y) G_0(x,y) \eqend{.}
\end{equations}

For the $\chi_1^\mu \chi_1^\nu$ correction we proceed similarly, using the path-ordering
\begin{equations}
\expect{ \mathcal{P} h_{\alpha\beta}(\tau') h_{\gamma\delta}(\tau'') } &= \Theta(\tau'-\tau'') \expect{ h_{\alpha\beta}(\tau') h_{\gamma\delta}(\tau'') } + \Theta(\tau''-\tau') \expect{ h_{\gamma\delta}(\tau'') h_{\alpha\beta}(\tau') } \eqend{,} \\
\expect{ \mathcal{P} h_{\alpha\beta}(\tau') h_{\gamma\delta}(x) } &= \expect{ h_{\alpha\beta}(\tau') h_{\gamma\delta}(x) } \eqend{,} \\
\expect{ \mathcal{P} h_{\alpha\beta}(x) h_{\gamma\delta}(\tau') } &= \expect{ h_{\gamma\delta}(\tau') h_{\alpha\beta}(x) } \eqend{.}
\end{equations}
Passing to Fourier space, using equation~\eqref{calculation_prelim_gravprop} and the fact that massless tadpoles (i.e., an expectation value of two gravitons at the same point) vanish in dimensional regularisation, we then obtain
\begin{splitequation}
\expect{ \chi_1^\mu(1) \chi_1^\nu(1) } &= \frac{1}{2} \sum_{i=1}^5 g_i (x-y)^\alpha (x-y)^\beta (x-y)^\rho \eta^{\gamma(\mu} \eta^{\nu)\delta} \\
&\qquad\times \int_0^1 (1-\tau') \int \frac{1}{k^2 - \mathi 0} \left[ k_\delta T^{(i)}_{\alpha\beta\gamma\rho}(k) - 2 k_\alpha T^{(i)}_{\beta\gamma\delta\rho}(k) \right] \mathe^{- \mathi k (x-y) \tau'} \frac{\total^n k}{(2\pi)^n} \total \tau' \\
&- \frac{\mathi}{4} \sum_{i=1}^5 g_i \int_0^1 (1-\tau') \int_0^{\tau'} (1-\tau'') (x-y)^\alpha (x-y)^\beta (x-y)^\gamma (x-y)^\delta \eta^{\mu\rho} \eta^{\nu\sigma} \\
&\qquad\times \int \left[ k_\rho k_\sigma T^{(i)}_{\alpha\beta\gamma\delta}(k) - 2 k_\alpha k_\sigma T^{(i)}_{\beta\rho\gamma\delta}(k) - 2 k_\rho k_\gamma T^{(i)}_{\alpha\beta\delta\sigma}(k) + 4 k_\alpha k_\gamma T^{(i)}_{\beta\rho\delta\sigma}(k) \right] \\
&\qquad\qquad\times \frac{1}{k^2 - \mathi 0} \mathe^{- \mathi k (x-y) (\tau'-\tau'')} \frac{\total^n k}{(2\pi)^n} \total \tau'' \total \tau' \\
&- \frac{\mathi}{4} \sum_{i=1}^5 g_i \int_0^1 (1-\tau') \int_{\tau'}^1 (1-\tau'') (x-y)^\alpha (x-y)^\beta (x-y)^\gamma (x-y)^\delta \eta^{\mu\rho} \eta^{\nu\sigma} \\
&\qquad\times \int \left[ k_\rho k_\sigma T^{(i)}_{\gamma\delta\alpha\beta}(k) - 2 k_\alpha k_\sigma T^{(i)}_{\gamma\delta\beta\rho}(k) - 2 k_\rho k_\gamma T^{(i)}_{\delta\sigma\alpha\beta}(k) + 4 k_\alpha k_\gamma T^{(i)}_{\delta\sigma\beta\rho}(k) \right] \\
&\qquad\qquad\times \frac{1}{k^2 - \mathi 0} \mathe^{\mathi k (x-y) (\tau'-\tau'')} \frac{\total^n k}{(2\pi)^n} \total \tau'' \total \tau' \eqend{.}
\end{splitequation}
We disentangle the $\tau$ and $k$ integrations by the rescalings $k \to k/\tau'$, $k \to k/(\tau'-\tau'')$ and $k \to k/(\tau'' - \tau')$, and perform the $\tau$ integrals to obtain
\begin{splitequation}
\expect{ \chi_1^\mu(1) \chi_1^\nu(1) } &= \frac{1}{2 (n-3)(n-2)} \sum_{i=1}^5 g_i (x-y)^\alpha (x-y)^\beta (x-y)^\rho \eta^{\gamma(\mu} \eta^{\nu)\delta} \\
&\qquad\times \int \frac{1}{k^2 - \mathi 0} \left[ k_\delta T^{(i)}_{\alpha\beta\gamma\rho}(k) - 2 k_\alpha T^{(i)}_{\beta\gamma\delta\rho}(k) \right] \mathe^{- \mathi k (x-y)} \frac{\total^n k}{(2\pi)^n} \\
&+ \frac{\mathi}{2 (n-1) (n-2) (n-4)} \sum_{i=1}^5 g_i (x-y)^\alpha (x-y)^\beta (x-y)^\gamma (x-y)^\delta \eta^{\mu\rho} \eta^{\nu\sigma} \\
&\qquad\times \int \left[ k_\rho k_\sigma T^{(i)}_{\alpha\beta\gamma\delta}(k) - 2 k_\alpha k_\sigma T^{(i)}_{\beta\rho\gamma\delta}(k) - 2 k_\rho k_\gamma T^{(i)}_{\alpha\beta\delta\sigma}(k) + 4 k_\alpha k_\gamma T^{(i)}_{\beta\rho\delta\sigma}(k) \right] \\
&\qquad\qquad\times \frac{1}{k^2 - \mathi 0} \mathe^{- \mathi k (x-y)} \frac{\total^n k}{(2\pi)^n} \eqend{,}
\end{splitequation}
using also that the tensor structures~\eqref{calculation_prelim_gravprop_tensors} are symmetric to simplify the result. The $k$ integrals can be evaluated in the same manner as for the $\chi_2^\mu$ correction, using in addition that
\begin{equation}
[ (x-y)^2 ]^2 \partial^\mu \partial^\nu G_0(x,y) \partial_\mu \partial_\nu G_0(x,y) = n (n-1) (n-2)^2 G_0(x,y) G_0(x,y) \eqend{,}
\end{equation}
and we obtain for the different tensor structures
\begin{equations}
\mathcal{G}_0^{\text{SG},11,(1)}(x,y) &= - \mathi \kappa^2 \frac{(n-2)^2}{2 (n-4)} G_0(x,y) G_0(x,y) \eqend{,} \\
\mathcal{G}_0^{\text{SG},11,(2)}(x,y) &= \mathi \kappa^2 \frac{(n-2)^2}{2 (n-3) (n-4)} G_0(x,y) G_0(x,y) \eqend{,} \\
\mathcal{G}_0^{\text{SG},11,(3)}(x,y) &= - \mathi \kappa^2 \frac{(n-1) (n-2)}{8 (n-3)} G_0(x,y) G_0(x,y) \eqend{,} \\
\mathcal{G}_0^{\text{SG},11,(4)}(x,y) &= \mathi \kappa^2 \frac{(n-1) (n-2)}{4} G_0(x,y) G_0(x,y) \eqend{,} \\
\mathcal{G}_0^{\text{SG},11,(5)}(x,y) &= \mathi \kappa^2 \frac{n (n-1) (n-2)^2}{4} G_0(x,y) G_0(x,y) \eqend{.}
\end{equations}

\subsection{Renormalising the geodesic corrections}
\label{sec_calculation_renorm}

Summing up the geodesic corrections, we obtain
\begin{equation}
\mathcal{G}_0^{\text{G},(i)}(x,y) = \mathcal{G}_0^{\text{FG},0,(i)}(x,y) + \mathcal{G}_0^{\text{FG},\tau,(i)}(x,y) + \mathcal{G}_0^{\text{SG},2,(i)}(x,y) + \mathcal{G}_0^{\text{SG},11,(i)}(x,y)
\end{equation}
with (up to terms which vanish as $n \to 4$)
\begin{splitequation}
\mathcal{G}_0^{\text{G},(1)}(x,y) &= \mathi \kappa^2 \bigg[ \frac{n-2}{4} \left( \frac{\Gamma(3-n) \Gamma(3-n)}{\Gamma(5-2n)} + \frac{3n^2-15n+20}{(n-3) (n-4)} \right) \\
&\qquad\quad+ \xi \left( \frac{n}{2} \frac{\Gamma(2-n) \Gamma(3-n)}{\Gamma(4-2n)} - 1 \right) \bigg] G_0(x,y) G_0(x,y) \\
&= \mathi \kappa^2 \bigg[ \frac{10+48\xi}{n-4} + \frac{31}{2} + 91 \xi + \left( \frac{17}{4} - \pi^2 + 4 \xi - 8 \pi^2 \xi \right) (n-4) \bigg] G_0(x,y) G_0(x,y)\eqend{,}
\end{splitequation}
\begin{splitequation}
\mathcal{G}_0^{\text{G},(2)}(x,y) &= \mathi \kappa^2 \frac{n-2}{8} \bigg[ n [ (n-2) - 4 (n-1) \xi ] \frac{\Gamma(2-n) \Gamma(3-n)}{\Gamma(5-2n)} + \frac{n^3-n^2-20n+40}{(n-3) (n-4)} \bigg] G_0(x,y) G_0(x,y) \\
&= \mathi \kappa^2 \bigg[ \frac{-10+72\xi}{n-4} - 25 + 162 \xi + \left( - \frac{45}{4} + 2 \pi^2 + 52 \xi - 12 \pi^2 \xi \right) (n-4) \bigg] G_0(x,y) G_0(x,y) \eqend{,}
\end{splitequation}
\begin{splitequation}
\mathcal{G}_0^{\text{G},(3)}(x,y) &= - \mathi \kappa^2 \bigg[ \frac{(n-2)}{8 (n-3)} (2n^2-13n+19) - \frac{(n-1)}{2} \xi \bigg] G_0(x,y) G_0(x,y) \\
&= \mathi \kappa^2 \bigg[ \frac{1}{4} + \frac{3}{2} \xi + \left( - \frac{7}{8} + \frac{1}{2} \xi \right) (n-4) \bigg] G_0(x,y) G_0(x,y) \eqend{,}
\end{splitequation}
\begin{equation}
\mathcal{G}_0^{\text{G},(4)}(x,y) = \mathi \kappa^2 \frac{(n-2) (5-n)}{4} G_0(x,y) G_0(x,y) \eqend{,}
\end{equation}
and
\begin{equation}
\mathcal{G}_0^{\text{G},(5)}(x,y) = \mathi \kappa^2 \frac{(n-2) (4n^3-12n^2+5n+7)}{16} G_0(x,y) G_0(x,y) \eqend{.}
\end{equation}
Using the explicit form of the massless scalar propagator~\eqref{calculation_prelim_g0}, it follows that
\begin{splitequation}
\label{calculation_renorm_geodesic}
\mathcal{G}_0^\text{G}(x,y) &= \sum_{i=1}^5 g_i \mathcal{G}_0^{\text{G},(i)}(x,y) \\
&= - \mathi \kappa^2 c_n^2 \bigg[ 4 \frac{5+6\xi}{n-4} + \frac{71}{2} - 35 \xi - 45 (\alpha-1) + \frac{91}{8} (\alpha-1) \beta^2 + \frac{1+6\xi}{4} (2+\beta) \\
&\qquad\quad- \frac{91}{4} (2+\beta)^2 + (n-4) \bigg[ \frac{11}{2} - 3 \pi^2 + 15 \xi + 4 \pi^2 \xi - \frac{147}{2} (\alpha-1) \\
&\qquad\quad+ \frac{293}{16} (\alpha-1) \beta^2 - \frac{4+\xi}{4} (2+\beta) - \frac{495}{16} (2+\beta)^2 \bigg] \bigg] \left[ (x-y)^2 \right]^{2-n} \eqend{.}
\end{splitequation}

As already noted in section~\ref{sec_calculation_firstgeodesic}, because of the divergent prefactor the result is divergent even for separated points $x \neq y$, and cannot be renormalised using only the higher-derivative counterterm $\delta_{(\partial^2 \phi)^2}$. To renormalise the above result, we borrow ideas from the renormalisation of Wilson loops in non-Abelian gauge theories~\cite{gervaisneveu1980,dotsenkovergeles1980,brandtnerisato1981} and post-Newtonian point particle dynamics in General Relativity~\cite{blanchetdamourespositofarese2004,blanchetetal2005,jaranowskischaefer2013,bernardetal2015}. In the case of Wilson loops on a smooth contour, there is an overall divergent factor proportional to the length of the contour, which can be renormalised by rewriting the path-ordered exponential of the gauge field as a one-dimensional fermion ``living on the contour'' coupled to the gauge field, and then performing the usual wave function and mass renormalisation for the fermion. In the gravitational case, the equations of motion for point particles contain divergences at higher orders in the post-Newtonian expansion, which can be renormalised by a (formally divergent) shift of the world lines of the particles proportional to the acceleration of the world line itself. Combining both ideas, we perform a ``wave function renormalisation'' of the geodesic $\chi^\mu(\tau)$, of the form
\begin{equation}
\label{calculation_renorm_chi}
\chi^\mu(\tau) \to Z_\chi \chi^\mu(\tau) = \chi^\mu(\tau) + \delta Z_\chi \chi^\mu(\tau) \eqend{.}
\end{equation}
Since we want $\delta Z_\chi = \bigo{\kappa^2}$, for dimensional reasons it must be proportional to $\ell^{-2}$, where $\ell$ is the length of the geodesic. In contrast to the situation for non-Abelian gauge theories where the counterterm is proportional to $\ell$, this means that the renormalisation itself is divergent as $\ell \to 0$. While this introduces a certain arbitrariness in intermediate steps, the final result nevertheless is a well-defined distribution. Noting that $x^\mu = \chi^\mu(0)$, this ``wave function renormalisation'' leads to an additional contribution to the two-point function at fixed geodesic distance of the form
\begin{splitequation}
\label{calculation_renorm_counterterm}
\mathcal{G}_0^\text{CT}(x,y) &= - \mathi \delta Z_\chi y^\mu \expect{ \phi(x) \partial_\mu \phi(y) } - \mathi \delta Z_\chi x^\mu \expect{ \partial_\mu \phi(x) \phi(y) } \\
&= \zeta \frac{\kappa^2}{(x-y)^2} (x-y)^\mu \partial_\mu G_0(x,y) = \mathi (n-2) c_n \zeta \kappa^2 \left[ (x-y)^2 \right]^{-\frac{n}{2}} \eqend{,}
\end{splitequation}
where we have written
\begin{equation}
\delta Z_\chi = \zeta \frac{\kappa^2}{\ell^2} + \bigo{\kappa^4} = \zeta \frac{\kappa^2}{(x-y)^2} + \bigo{\kappa^4}
\end{equation}
with a constant $\zeta$, and used the explicit form~\eqref{calculation_prelim_g0} of the massless scalar propagator.

Because of translation invariance, we can set $y = 0$. To shorten intermediate steps, let us write the result~\eqref{calculation_renorm_geodesic} in the form
\begin{equation}
\mathcal{G}_0^\text{G}(x,0) = - \mathi \kappa^2 c_n^2 \left[ \frac{A}{n-4} + B + C (n-4) \right] (x^2)^{2-n}
\end{equation}
with the constants
\begin{equations}[calculation_renorm_abc]
A &= 4 (5+6\xi) \eqend{,} \\
B &= \frac{71}{2} - 35 \xi - 45 (\alpha-1) + \frac{91}{8} (\alpha-1) \beta^2 + \frac{1+6\xi}{4} (2+\beta) - \frac{91}{4} (2+\beta)^2 \eqend{,}
\end{equations}
and $C$ (whose precise expression is unimportant), of which only $A$ is gauge-independent. We then proceed similarly to the field-theoretic corrections, using equation~\eqref{appendix_feynman_x2_dalembert_p} to extract a d'Alembertian operator and adding an intelligent zero to obtain
\begin{splitequation}
(x^2)^{2-n} &= \frac{1}{2 (n-3) (n-4)} \partial^2 (x^2)^{3-n} \\
&= \frac{1}{2 (n-3) (n-4)} \partial^2 \left[ (x^2)^{3-n} - \mu^{n-4} (x^2)^{1-\frac{n}{2}} \right] + \frac{\mathi}{2 c_n (n-3) (n-4)} \mu^{n-4} \delta^n(x) \eqend{.}
\end{splitequation}
Expanding around $n = 4$, this gives
\begin{splitequation}
- \mathi c_n^2 (x^2)^{2-n} &= H^{(1)}_0(x;\mu) + (n-4) \left( 1 + 2 \ln \bar{\mu} \right) H^{(1)}_0(x;\mu) - \frac{3}{4} (n-4) H^{(2)}_0(x;\mu) \\
&\quad+ \frac{1}{16 \pi^2} \left[ \frac{2}{n-4} + 2 \ln \bar{\mu} + (n-4) \left( \frac{\pi^2}{24} + 1 + \ln^2 \bar{\mu} \right) \right] \delta^n(x) + \bigo{(n-4)^2} \eqend{,}
\end{splitequation}
with the distributions $H^{(k)}_0$ and the parameter $\bar{\mu}$ defined in equations~\eqref{calculation_selfenergy_h0_def} and~\eqref{calculation_selfenergy_mubar}, respectively, and thus
\begin{splitequation}
\label{calculation_renorm_geodesic2}
\mathcal{G}_0^\text{G}(x,0) &= \frac{A}{n-4} \kappa^2 H^{(1)}_0(x;\mu) + \left( A + B + 2 A \ln \bar{\mu} \right) \kappa^2 H^{(1)}_0(x;\mu) - \frac{3}{4} A \kappa^2 H^{(2)}_0(x;\mu) \\
&\quad+ \frac{\kappa^2}{16 \pi^2} \bigg[ \frac{2 A}{(n-4)^2} + \frac{2}{n-4} \left( A \ln \bar{\mu} + B \right) \\
&\qquad\qquad+ A \left( \frac{\pi^2}{24} + 1 + \ln^2 \bar{\mu} \right) + 2 B \ln \bar{\mu} + 2 C \bigg] \delta^n(x) + \bigo{n-4} \eqend{.}
\end{splitequation}
If we try to perform the same procedure with the counterterm~\eqref{calculation_renorm_counterterm}, we immediately run into problems because the formula~\eqref{appendix_feynman_x2_dalembert_p} is ill-defined for $(x^2)^{-\frac{n}{2}}$, on account of the later not being a well-defined distribution in \emph{any} dimension $n$. To ameliorate the problem, we write
\begin{splitequation}
\label{calculation_renorm_geodesicct}
(x^2)^{-\frac{n}{2}} &= \mu^{- 2 \delta (n-4)} (x^2)^{- \frac{n}{2} - \delta (n-4)} \\
&\quad\times \bigg[ 1 + \delta (n-4) \ln (\mu^2 x^2) + \frac{\delta^2}{2} (n-4)^2 \ln^2 (\mu^2 x^2) + \bigo{(n-4)^3} \bigg]
\end{splitequation}
for some $\delta$ and take the limit $\delta \to 0$ in the end, which should be possible for all non-local terms. As long as $\delta > 0$, every term of the expansion is a well-defined distribution for suitable (small) dimension $n$, and we can use the same techniques for renormalisation. Generalising equation~\eqref{appendix_feynman_x2_dalembert_p} to include logarithms, we calculate
\begin{splitequation}
&\partial^2 \left[ (x^2)^{1-p} \ln^q (\mu^2 x^2) \right] \\
&\quad= 2 \left[ (1-p) (n-2p) \ln^2 (\mu^2 x^2) + q (n+2-4p) \ln (\mu^2 x^2) + 2 q (q-1) \right] (x^2)^{-p} \ln^{q-2} (\mu^2 x^2)
\end{splitequation}
and from this
\begin{equations}
(x^2)^{-p} &= \frac{1}{2 (1-p) (n-2p)} \partial^2 (x^2)^{1-p} \eqend{,} \\
(x^2)^{-p} \ln (\mu^2 x^2) &= \frac{1}{2 (1-p) (n-2p)} \partial^2 \left[ (x^2)^{1-p} \left( \ln (\mu^2 x^2) - \frac{n+2-4p}{(1-p) (n-2p)} \right) \right] \eqend{,} \\
\begin{split}
(x^2)^{-p} \ln^2 (\mu^2 x^2) &= \frac{1}{2 (1-p) (n-2p)} \partial^2 \bigg[ (x^2)^{1-p} \bigg( \ln^2 (\mu^2 x^2) - \frac{2 (n+2-4p)}{(1-p) (n-2p)} \ln (\mu^2 x^2) \\
&\hspace{10em}+ \frac{2 (n^2+2n+4 - 6(n+2)p + 12p^2)}{(1-p)^2 (n-2p)^2} \bigg) \bigg] \eqend{.}
\end{split}
\end{equations}
It follows that
\begin{splitequation}
(x^2)^{-\frac{n}{2}} &= \mu^{- 2 \delta (n-4)} \partial^2 \bigg[ (x^2)^{1-\frac{n}{2} - \delta (n-4)} \bigg( \frac{1}{16 \delta} \left( 12 \frac{1}{n-4} - 2 (3+2\delta) + (3+4\delta) (n-4) \right) \\
&\qquad+ \left( \frac{1}{2} - \frac{1}{4} (1+\delta) (n-4) \right) \ln (\mu^2 x^2) + \frac{1}{8} \delta (n-4) \ln^2 (\mu^2 x^2) \bigg) \bigg] + \bigo{(n-4)^2} \\
&= \partial^2 \bigg[ (x^2)^{1-\frac{n}{2}} \bigg( \frac{1}{16 \delta} \left( 12 \frac{1}{n-4} - 2 (3+2\delta) + (3+4\delta) (n-4) \right) \\
&\qquad- \left( \frac{1}{4} - \frac{1}{8} (n-4) \right) \ln (\mu^2 x^2) \bigg) \bigg] + \bigo{(n-4)^2} \eqend{.}
\end{splitequation}
From the explicit expression of the massless propagator~\eqref{calculation_prelim_g0} we have
\begin{equation}
\partial^2 (x^2)^{1-\frac{n}{2}} = \frac{\mathi}{c_n} \delta^n(x) \eqend{,}
\end{equation}
and writing
\begin{equation}
\zeta = \frac{\zeta_0}{n-4} + \zeta_1 + (n-4) \zeta_2 \eqend{,}
\end{equation}
we obtain
\begin{splitequation}
\label{calculation_renorm_counterterm2}
\mathcal{G}_0^\text{CT}(x,0) &= - \frac{\kappa^2}{4} \left[ \frac{6}{\delta} \left( \frac{1}{(n-4)^2} \zeta_0 + \frac{1}{n-4} \zeta_1 + \zeta_2 \right) - \frac{2}{n-4} \zeta_0 + \zeta_0 - 2 \zeta_1 \right] \delta^n(x) \\
&\quad- 4 \pi^2 \kappa^2 \left[ \frac{2}{n-4} \zeta_0 + 2 \zeta_0 \left( \ln \bar{\mu} + 1 \right) + 2 \zeta_1 \right] H^{(1)}_0(x;\mu) \\
&\quad+ 4 \pi^2 \kappa^2 \zeta_0 H^{(2)}_0(x;\mu) + \bigo{n-4} \eqend{,}
\end{splitequation}
with the distributions $H^{(k)}_0$ and the parameter $\bar{\mu}$ defined in equations~\eqref{calculation_selfenergy_h0_def} and~\eqref{calculation_selfenergy_mubar}, respectively. Note that while all non-local terms [containing $H^{(k)}_0(x;\mu)$] are independent of the arbitrary parameter $\delta$, the limit $\delta \to 0$ cannot be taken for the local term proportional to $\delta^n(x)$. This is of course a consequence of the fact that the original counterterm $(x^2)^{-\frac{n}{2}}$ is not a well-defined distribution in any dimension $n$, but we see that the problem only appears at coincidence, as already alluded to previously.

In order for the sum $\mathcal{G}_0^\text{G}(x,0) + \mathcal{G}_0^\text{CT}(x,0)$ to have a finite coefficient in front of $H^{(1)}_0(x;\mu)$, comparing equations~\eqref{calculation_renorm_geodesic2} and~\eqref{calculation_renorm_counterterm2} we need to take
\begin{equation}
\zeta_0 = \frac{A}{8 \pi^2} \eqend{.}
\end{equation}
In order not to have spurious factors of $\ln \mu$, we furthermore have to set
\begin{equation}
\label{calculation_renorm_zeta1}
\zeta_1 = \frac{B + A \ln \bar{\mu}}{8 \pi^2} + \zeta^\text{fin}(\mu) \eqend{,}
\end{equation}
and obtain
\begin{splitequation}
\mathcal{G}_0^\text{CT}(x,0) + \mathcal{G}_0^\text{G}(x,0) &= - \frac{\kappa^2}{16 \pi^2} \frac{3-2\delta}{\delta} \left[ \frac{A}{(n-4)^2} + \frac{1}{n-4} \left( B + A \ln \bar{\mu} + 8 \pi^2 \zeta^\text{fin}(\mu) \right) + 8 \pi^2 \zeta_2 \right] \delta^n(x) \\
&\quad+ \frac{\kappa^2}{16 \pi^2} \left[ \frac{1}{n-4} \left( A - 16 \pi^2 \zeta^\text{fin}(\mu) \right) + A \frac{\pi^2}{24} + \frac{A}{2} + B + 2 C + A \ln^2 \bar{\mu} + (A+2B) \ln \bar{\mu} + 8 \pi^2 \zeta^\text{fin}(\mu) - 16 \pi^2 \zeta_2 \right] \delta^n(x) \\
&\quad- \frac{A}{4} \kappa^2 H^{(2)}_0(x;\mu) - 8 \pi^2 \kappa^2 \zeta^\text{fin}(\mu) H^{(1)}_0(x;\mu) + \bigo{n-4} \eqend{.}
\end{splitequation}
Finally, the contribution proportional to $\delta^n(x)$ can be absorbed using a higher-derivative counterterm of the form~\eqref{calculation_selfenergy_deltact}, taking (\emph{in addition} to the field-theoretic correction)
\begin{splitequation}
\label{calculation_renorm_deltad2phi2}
\delta_{(\partial^2 \phi)^2} &= \frac{\kappa^2}{16 \pi^2} \frac{3-2\delta}{2 \delta} \left[ \frac{A}{(n-4)^2} + \frac{1}{n-4} \left( B + A \ln \bar{\mu} + 8 \pi^2 \zeta^\text{fin}(\mu) \right) + 8 \pi^2 \zeta_2 \right] \\
&\quad- \frac{\kappa^2}{32 \pi^2} \left[ \frac{1}{n-4} \left( A - 16 \pi^2 \zeta^\text{fin}(\mu) \right) + A \frac{\pi^2}{24} + \frac{A}{2} + B + 2 C + A \ln^2 \bar{\mu} + (A+2B) \ln \bar{\mu} + 8 \pi^2 \zeta^\text{fin}(\mu) - 16 \pi^2 \zeta_2 \right] \\
&\quad+ \delta^\text{fin}_{(\partial^2 \phi)^2}(\mu) \eqend{,}
\end{splitequation}
such that the final renormalised result for the geodesic corrections reads [inserting the constant $A$~\eqref{calculation_renorm_abc}]
\begin{splitequation}
- (5+6\xi) \kappa^2 H^{(2)}_0(x;\mu) - 8 \pi^2 \kappa^2 \zeta^\text{fin}(\mu) H^{(1)}_0(x;\mu) - 2 \delta^\text{fin}_{(\partial^2 \phi)^2}(\mu) \delta^n(x) \eqend{.}
\end{splitequation}
For the special value $\delta = 3/2$, only single poles in $(n-4)$ appear. While a decomposition of the form~\eqref{calculation_renorm_geodesicct} would be valid for arbitrary $\delta$ if the left-hand side were a well-defined distribution for some $n$, and the result correspondingly independent of $\delta$, this is not the case here. Finally, from the $\mu$-independence of the bare counterterms $\zeta$ and $\delta_{(\partial^2 \phi)^2}$ and equations~\eqref{calculation_renorm_zeta1}, \eqref{calculation_renorm_deltad2phi2} and~\eqref{calculation_renorm_abc}, we calculate
\begin{equations}
\mu \frac{\total}{\total \mu} \zeta^\text{fin}(\mu) &= - \frac{5+6\xi}{2 \pi^2} \eqend{,} \\
\begin{split}
\mu \frac{\total}{\total \mu} \delta^\text{fin}_{(\partial^2 \phi)^2}(\mu) &= \frac{\kappa^2}{4 \pi^2} \bigg[ (5+6\xi) \left( \frac{1}{n-4} + \ln \bar{\mu} \right) + \frac{71}{8} - \frac{35}{4} \xi - \frac{45}{4} (\alpha-1) \\
&\qquad+ \frac{91}{32} (\alpha-1) \beta^2 + \frac{1+6\xi}{16} (2+\beta) - \frac{91}{16} (2+\beta)^2 \bigg] \eqend{.}
\end{split}
\end{equations}
Because of the term proportional to $(n-4)^{-1}$, this is intrinsically ill-defined; of course, this is just another consequence of the original geodesic counterterm $(x^2)^{- \frac{n}{2}}$ not being a well-defined distribution in any dimension $n$.

\section{Results}
\label{sec_results}

Adding all terms together and performing a further finite renormalisation for $\zeta^\text{fin}(\mu)$ to absorb the gauge-dependent field-theoretic corrections~\eqref{calculation_selfenergy_result}, the final result for the scalar two-point function at fixed geodesic distance $\ell^2 = (x-y)^2$ including one-loop graviton corrections reads
\begin{splitequation}
\label{results_2pf}
\mathcal{G}_0(x,y) &= G_0(x,y) - (5+6\xi) \kappa^2 H^{(2)}_0(x-y;\mu) - 8 \pi^2 \kappa^2 \zeta^\text{fin}(\mu) H^{(1)}_0(x-y;\mu) \\
&\quad- 2 \delta^\text{fin}_{(\partial^2 \phi)^2}(\mu) \delta^n(x-y) \eqend{,}
\end{splitequation}
where the distributions $H^{(k)}_0(x;\mu)$ are given by~\eqref{calculation_selfenergy_h0_def}
\begin{equation}
H^{(k)}_0(x;\mu) = \frac{\mathi}{64 \pi^4} \partial^2 \left[ \frac{\ln^k( \mu^2 x^2 )}{x^2} \right] \eqend{,}
\end{equation}
and the finite parts of counterterms $\zeta^\text{fin}(\mu)$ and $\delta^\text{fin}_{(\partial^2 \phi)^2}(\mu)$ depend on the renormalisation scale $\mu$ according to
\begin{equations}[results_mu_scaling]
\mu \frac{\total}{\total \mu} \zeta^\text{fin}(\mu) &= - \frac{5+6\xi}{2 \pi^2} \eqend{,} \\
\begin{split}
\mu \frac{\total}{\total \mu} \delta^\text{fin}_{(\partial^2 \phi)^2}(\mu) &= \frac{\kappa^2}{4 \pi^2} \bigg[ (5+6\xi) \left( \frac{1}{n-4} + \ln \bar{\mu} \right) + \frac{71}{8} - 9 \xi - \frac{177}{16} (\alpha-1) \\
&\qquad+ \frac{179}{64} (\alpha-1) \beta^2 + \frac{1+12\xi}{16} (2+\beta) - \frac{361}{64} (2+\beta)^2 \bigg] \eqend{.}
\end{split}
\end{equations}
We see that the result is intrinsically ill-defined at coincidence since the scaling of the finite parts with the renormalisation scale is infinite in $n = 4$ dimensions, but for $x \neq y$ we have obtained a finite and gauge-independent result. Since we have
\begin{equations}
\mu \frac{\total}{\total \mu} H^{(2)}_0(x;\mu) &= 4 H^{(1)}_0(x;\mu) \eqend{,} \\
\mu \frac{\total}{\total \mu} H^{(1)}_0(x;\mu) &= - \frac{1}{8 \pi^2} \delta^n(x) \eqend{,}
\end{equations}
we can alternatively define the scaling of $\zeta^\text{fin}(\mu)$ and $\delta^\text{fin}_{(\partial^2 \phi)^2}(\mu)$ by imposing the $\mu$-independence of $\mathcal{G}_0(x,y)$. This results in
\begin{equation}
\label{results_scaling_a}
\mu \frac{\total}{\total \mu} \zeta^\text{fin}(\mu) = - \frac{5+6\xi}{2 \pi^2} \quad \Rightarrow \quad \zeta^\text{fin}(\mu) = \zeta^\text{fin}(\mu_0) - \frac{5+6\xi}{2 \pi^2} \ln \frac{\mu}{\mu_0}
\end{equation}
and
\begin{splitequation}
\label{results_scaling_b}
\mu \frac{\total}{\total \mu} \delta^\text{fin}_{(\partial^2 \phi)^2}(\mu) &= \frac{1}{2} \kappa^2 \zeta^\text{fin}(\mu) = \frac{1}{2} \kappa^2 \zeta^\text{fin}(\mu_0) - \kappa^2 \frac{5+6\xi}{4 \pi^2} \ln \frac{\mu}{\mu_0} \\
\quad \Rightarrow \quad \delta^\text{fin}_{(\partial^2 \phi)^2}(\mu) &= \delta^\text{fin}_{(\partial^2 \phi)^2}(\mu_0) + \frac{1}{2} \kappa^2 \zeta^\text{fin}(\mu_0) \ln \frac{\mu}{\mu_0} - \kappa^2 \frac{5+6\xi}{8 \pi^2} \ln^2 \frac{\mu}{\mu_0} \eqend{,}
\end{splitequation}
which for $\zeta^\text{fin}(\mu)$ is identical to~\eqref{results_mu_scaling}.

The expression~\eqref{results_2pf} together with the scalings~\eqref{results_scaling_a} and~\eqref{results_scaling_b} give a finite and gauge-invariant scalar two-point function including one-loop graviton corrections, which is the main result of this paper. It can be clearly seen that the double logarithm inside $H^{(2)}$, which is highly unusual for a one-loop result, arises from the double pole $\sim (n-4)^{-2}$ at coincidence. This in turn comes from the restriction of the distribution-valued graviton operator $h_{\mu\nu}$ to the geodesic, whereby new UV divergences appear. However, the gauge dependence of intermediate steps and the fact that the scaling of $\delta^\text{fin}_{(\partial^2 \phi)^2}(\mu)$ need to be imposed by hand are puzzling. While we do not have any explanation for the second part, we can explain the first, which is done in the next subsection.

\subsection{Gauge dependence}

To study the gauge dependence of our correlation function at fixed geodesic distance, it suffices to work to first order. To this order, the infinitesimal coordinate change $x^\mu \to x^\mu - \kappa \xi^\mu$ induces the gauge transformation
\begin{equations}
\delta_\xi h_{\mu\nu} &= \mathcal{L}_\xi \eta_{\mu\nu} + \bigo{\kappa} = \partial_\mu \xi_\nu + \partial_\nu \xi_\mu + \bigo{\kappa} \eqend{,} \\
\delta_\xi \phi &= \mathcal{L}_\xi \phi = \kappa \xi^\mu \partial_\mu \phi \eqend{,}
\end{equations}
where $\mathcal{L}_\xi$ is the Lie derivative with respect to $\xi^\mu$. We then calculate for the gauge variation of the perturbed geodesic coordinate~\eqref{calculation_prelim_firstordergeodesic}
\begin{splitequation}
\delta_\xi \chi_1^\mu(\tau) &= \frac{1}{2} \left( \partial^\mu \xi_\rho(x) + \partial_\rho \xi^\mu(x) \right) (x-y)^\rho \tau \\
&\quad- \int_0^\tau (\tau - \tau') (x-y)^\alpha (x-y)^\beta \partial_\alpha \partial_\beta \xi^\mu(\tau') \total \tau' + \bigo{\kappa} \eqend{.}
\end{splitequation}
Since
\begin{equations}
(x-y)^\alpha (x-y)^\beta \partial_\alpha \partial_\beta \xi^\mu(\tau') &= \partial^2_{\tau'} \xi^\mu(\tau') \eqend{,} \\
(x-y)^\rho \partial_\rho \xi^\mu(\tau') &= - \partial_{\tau'} \xi^\mu(\tau') \eqend{,}
\end{equations}
this reduces after integration by parts to
\begin{equation}
\label{results_gauge_deltachi}
\delta_\xi \chi_1^\mu(\tau) = \frac{1}{2} \left[ \partial^\mu \xi_\rho(x) - \partial_\rho \xi^\mu(x) \right] (x-y)^\rho \tau - \xi^\mu(\chi_0(\tau)) + \xi^\mu(x) + \bigo{\kappa} \eqend{.}
\end{equation}
However, this is not the full result for the gauge variation. For the initial direction of the geodesic~\eqref{calculation_prelim_geodesic_bdy2}, we fixed the bein components $v^a$, and subsequently choose symmetric gauge for the local Lorentz symmetry of the $n$-bein. To keep this gauge condition, we must compensate for the coordinate change by an explicit Lorentz transformation~\cite{vannieuwenhuizen1981,woodard_thesis,woodard1984}. Namely, the change of the $n$-bein under the above infinitesimal coordinate transformation is given by
\begin{splitequation}
\delta_\xi e_\mu{}^a &= \mathcal{L}_\xi e_\mu{}^a = \kappa \delta_\rho^a \partial_\mu \xi^\rho + \bigo{\kappa^2} \\
&= \frac{\kappa}{2} \delta_\rho^a \left( \partial_\mu \xi^\rho + \partial^\rho \xi_\mu \right) + \frac{\kappa}{2} \delta_\rho^a \left( \partial_\mu \xi^\rho - \partial^\rho \xi_\mu \right) + \bigo{\kappa^2} \eqend{.}
\end{splitequation}
The first, symmetric term is the one that would be obtained from the transformation of the explicit expansion~\eqref{appendix_metric_bein_expansion} of the $n$-bein in symmetric gauge. The second one can be cancelled by a compensating Lorentz transformation of the form
\begin{equation}
\delta_\omega e_\mu{}^a = e_\mu{}^b \left[ \Lambda_b{}^a(\omega) - \delta_b^a \right] = \frac{\kappa}{2} \eta_{bc} \omega^{ab} e_\mu{}^c + \bigo{\kappa^2}
\end{equation}
with the antisymmetric Lorentz parameter
\begin{equation}
\label{results_gauge_lorentz}
\omega^{ab} = e^a_\mu e^b_\nu \left( \partial^\mu \xi^\nu - \partial^\nu \xi^\mu \right) \eqend{.}
\end{equation}
Taking both together, we obtain the correct transformation
\begin{equation}
\left( \delta_\xi + \delta_\omega \right) e_\mu{}^a = \frac{\kappa}{2} \delta_\rho^a \left( \partial_\mu \xi^\rho + \partial^\rho \xi_\mu \right) + \bigo{\kappa^2} \eqend{.}
\end{equation}
Since we have used the explicit expansion of the $n$-bein in the gauge transformation of the geodesic coordinate~\eqref{results_gauge_deltachi}, which only takes into account the symmetric part, we need to perform the Lorentz transformation~\eqref{results_gauge_lorentz} on the initial direction vector $v^a$. This results in
\begin{equations}
\delta_\xi v^\mu &= \frac{\kappa}{2} \left( \partial^\mu \xi_\nu - \partial_\nu \xi^\mu \right) v^\nu + \bigo{\kappa^2} \eqend{,} \\
\delta_\xi \chi_0^\mu(\tau) &= - \frac{\kappa}{2} \left( \partial^\mu \xi_\nu(x) - \partial_\nu \xi^\mu(x) \right) (x-y)^\nu \tau + \bigo{\kappa^2} \eqend{,}
\end{equations}
and we finally have
\begin{equation}
\delta_\xi \left[ \chi_0^\mu(\tau) + \kappa \chi_1^\mu(\tau) \right] = - \kappa \xi^\mu(\chi_0(\tau)) + \kappa \xi^\mu(x) + \bigo{\kappa^2} \eqend{.}
\end{equation}
Of course, this result is also obtained by restricting to symmetric gauge transformations $\partial_\mu \xi_\rho = \partial_\rho \xi_\mu$ for which the compensating Lorentz transformation~\eqref{results_gauge_lorentz} vanishes.

Since
\begin{equation}
\phi(\chi(\tau)) = \phi(\chi_0(\tau)) + \kappa \chi_1^\mu(\tau) \partial_\mu \phi(\chi_0(\tau)) + \bigo{\kappa^2} \eqend{,}
\end{equation}
it follows that
\begin{equation}
\label{results_gauge_deltaphi}
\delta_\xi \phi(\chi(\tau)) = \kappa \xi^\mu(x) \partial^x_\mu \phi(x+v\tau) + \bigo{\kappa^2} \eqend{.}
\end{equation}
To obtain a fully invariant variable, we would need to multiply by $\sqrt{-g(x)}$, which to first order transforms as a total derivative,
\begin{splitequation}
\delta_\xi \left[ \sqrt{-g(x)} \phi(\chi(\tau)) \right] &= \delta_\xi \left[ \left( 1 + \frac{1}{2} \kappa h(x) \right) \phi(\chi(\tau)) \right] + \bigo{\kappa^2} \\
&= \kappa \partial^x_\mu \left[ \xi^\mu(x) \phi(x+v\tau) \right] + \bigo{\kappa^2} \eqend{,}
\end{splitequation}
and then integrate over $x$, assuming as usual that the gauge transformation is sufficiently fast decaying at infinity.

The correlation function~\eqref{results_2pf} that we calculated does not include the factors of $\sqrt{-g}$. However, since massless tadpoles vanish in dimensional regularisation, only its first-order expansion contributes, which is proportional to the trace $h$ of the metric perturbation. Since the trace of the graviton propagator~\eqref{calculation_prelim_trace} is proportional to the gauge parameter $\beta$, including the factors of $\sqrt{-g}$ can only contribute terms which contain at least one $\beta$; in particular, in the gauge $\beta = 0$ including factors of $\sqrt{-g}$ does not change the end result. However, even for $\beta = 0$ both the regularised result for the field-theoretic corrections~\eqref{calculation_selfenergy_regresult} and the geodesic corrections~\eqref{calculation_renorm_geodesic} depend on the other gauge parameter $\alpha$; for their sum we obtain
\begin{splitequation}
&\left[ \mathcal{G}^\text{FT,33}_0(x,y) + \mathcal{G}_0^\text{G}(x,y) \right]_{\beta = 0} \\
&= - \mathi \kappa^2 c_n^2 \left[ 4 \frac{5+6\xi}{n-4} - 55 - 34 \xi - \frac{183}{4} (\alpha-1) + \bigo{n-4} \right] \left[ (x-y)^2 \right]^{2-n} \eqend{.}
\end{splitequation}
While the direction of the geodesic is invariantly specified by fixing the bein components of the initial direction $v^a$, the starting point $x^\mu$ is not, and the correlation function is only invariant under those gauge transformations that leave $x^\mu$ unchanged, i.e., $\xi^\mu(x) = 0$, as can be seen directly from equation~\eqref{results_gauge_deltaphi}. This is then reflected in the regularised result; why the renormalisation of the geodesic~\eqref{calculation_renorm_chi} removes all gauge dependence is so far unclear and merits further investigation.

\section{Discussion}
\label{sec_discussion}

We have calculated the scalar two-point function at fixed geodesic distance in a flat-space background, including the one-loop effects of virtual gravitons. The result is given by equation~\eqref{results_2pf} together with the renormalisation group scalings~\eqref{results_scaling_a} and~\eqref{results_scaling_b}. By renormalising the novel UV divergences which appear in the correlation function using a ``wave function renormalisation'' of the geodesic itself, we have obtained a finite and gauge-invariant result in four dimensions, which to our knowledge is the first fully renormalised result of a correlation function at fixed geodesic distance in perturbative quantum gravity. It has some unusual features, which all ultimately stem from the additional UV divergences: a double logarithm appears already at the one-loop level, while in usual perturbative calculations only involving bulk integrals, at most $k$ logarithms can arise at the $k$-loop level, and correspondingly a dependence on the finite parts of higher-derivative counterterms even for separated points. This is in strong contrast to existing quantum gravitational results at one-loop order, e.g., the corrections to the Newtonian potential~\cite{schwinger1968,radkowski1970,satzmazzitellialvarez2005,parkwoodard2010,marunovicprokopec2011,marunovicprokopec2012,burnspilaftsis2015,froeb2016,donoghue1994a,muzinichvokos1995,hamberliu1995,akhundovbelluccishiekh1997,kirilinkhriplovich2002,donoghueetal2002,khriplovichkirilin2003,bjerrumbohrdonoghueholstein2003a,bjerrumbohrdonoghueholstein2003b,holsteinross2008}, where the result is unambiguous for separated points and the dependence on the finite parts of higher-derivative counterterms only appears at coincidence.

A related issue is the connection of the two-point function at fixed geodesic distance to results obtained using the S-matrix. Using our result~\eqref{results_2pf} to quantify quantum gravitational corrections to a scalar interaction potential, the double logarithm would generate corrections of the form $\kappa^2 r^{-3} \ln (\mu r)$ to the tree-level $1/r$ potential. However, using the inverse scattering method to reconstruct a non-relativistic scattering potential from the S-matrix element for the interaction of two scalars including graviton corrections, the one-loop correction is of the form $\kappa^2 r^{-3}$ (with the coefficient depending on the details of the interaction), which arises from a single logarithm. Certainly both approaches should be valid, but one has to find the precise connection between them (which might involve delicate cancellations), and to clarify their exact relation with actual experiments. Note that in the case of matter corrections, the terms in the quantum-corrected graviton two-point function which would give a contribution proportional to $\kappa^2 r^{-3} \ln (\mu r)$ are pure gauge, and when coupled to the conserved stress tensor of a point particle do not make a contribution to the Newton potential~\cite{duff1974,duffliu2000}.\footnote{I thank Michael Duff for bringing this point to my attention.} This has also implications for a generalisation of the present approach to inflationary spacetimes. It is known that loop corrections to correlation functions in inflation are infrared (IR) divergent~\cite{giddingssloth2011}. To obtain physical, IR finite correlation functions it has been suggested to take long wavelength fluctuations of the metric into account when defining distances, i.e., fixing the geodesic instead of the background distance~\cite{urakawatanaka2010b,gerstenlaueretal2011}. It then has been shown that IR divergences are indeed ameliorated by this approach, while it was tacitly assumed that the UV structure of the correlation functions is unchanged, and divergences can be renormalised in the usual way. The results of the present work show that one needs to be careful in making such an assumption, and indeed for our correlation functions it does not hold (we note that the concrete proposal of refs.~\cite{urakawatanaka2010b,gerstenlaueretal2011} is slightly different from ours).

Some points also merit further investigation from a more mathematical standpoint. Naturally, it will be important to see if the geodesic wave function renormalisation is also sufficient to achieve a finite result at higher loop orders. Another important issue concerns the gauge dependence: although the final renormalised result is gauge-independent, the regularised expression is not, with the gauge dependence being cancelled by the geodesic counterterm. While the dependence of counterterms on the gauge is not unusual, and the reason for the gauge dependence of the regulated expression has been explained, it is not clear whether this cancellation persists at higher orders, or for different correlation functions. For the usual perturbative calculations in gauge field theories, one can use the BRST formalism~\cite{becchietal1975} (or the extension due to Batalin and Vilkovisky~\cite{batalinvilkovisky1981,batalinvilkovisky1983,batalinvilkovisky1984} for open gauge algebras) to prove Ward identities to all orders in perturbation theory, which imply gauge independence for the correlation functions of BRST-invariant operators (possibly including quantum corrections to the BRST differential). If furthermore there exists a regulator which formally conserves BRST invariance, such as dimensional regularisation for most gauge theories including perturbative quantum gravity, already the regulated correlation functions are gauge-independent --- which is not the case in the calculation at hand. It remains to be seen if the theory (including fluctuations of the geodesic) can be reformulated in such a way that BRST invariance is manifest throughout all stages of the calculation, which would imply gauge independence of the final results to all orders in perturbation theory.

\ack
It is a pleasure to thank Atsushi Higuchi and Richard Woodard for discussions, in particular on gauge invariance and on renormalising the geodesic corrections, and Sohyun Park for discussions on refs.~\cite{borankahyapark2014,borankahyapark2017}. This work is part of a project that has received funding from the European Union’s Horizon 2020 research and innovation programme under the Marie Sk{\l}odowska-Curie grant agreement No. 702750 ``QLO-QG''.

\appendix

\section{Metric expansions}
\label{appendix_metric}

Writing a general metric $\tilde{g}_{\mu\nu}$ as background $g_{\mu\nu}$ plus perturbation $h_{\mu\nu}$, we obtain to first order in the perturbation
\begin{equations}[appendix_metric_expansion]
\tilde{g}_{\mu\nu} &= g_{\mu\nu} + \kappa h_{\mu\nu} \eqend{,} \\
\tilde{g}^{\mu\nu} &= g^{\mu\nu} - \kappa h^{\mu\nu} + \bigo{\kappa^2} \eqend{,} \\
\sqrt{-\tilde{g}} &= \sqrt{-g} \left( 1 + \frac{1}{2} \kappa h \right) + \bigo{\kappa^2} \eqend{,} \\
\tilde{\Gamma}^\alpha_{\beta\gamma} &= \Gamma^\alpha_{\beta\gamma} + \frac{1}{2} \kappa \left( \nabla_\beta h^\alpha_\gamma + \nabla_\gamma h^\alpha_\beta - \nabla^\alpha h_{\beta\gamma} \right) + \bigo{\kappa^2} \eqend{,} \\
\begin{split}
\tilde{R}_{\alpha\beta\gamma\delta} &= R_{\alpha\beta\gamma\delta} + \frac{1}{2} \kappa \left( \nabla_\gamma \nabla_{[\beta} h_{\alpha]\delta} - \nabla_\delta \nabla_{[\beta} h_{\alpha]\gamma} + \nabla_\alpha \nabla_{[\delta} h_{\gamma]\beta} - \nabla_\beta \nabla_{[\delta} h_{\gamma]\alpha} \right) \\
&\hspace{4em}- \frac{1}{2} \kappa \left( R_{\alpha\beta\mu[\gamma} h_{\delta]}^\mu + R_{\gamma\delta\mu[\alpha} h_{\beta]}^\mu \right) + \bigo{\kappa^2} \eqend{,}
\end{split} \\
\tilde{R}_{\alpha\beta} &= R_{\alpha\beta} + \kappa \nabla^\delta \nabla_{(\alpha} h_{\beta)\delta} - \frac{1}{2} \kappa \nabla^2 h_{\alpha\beta} - \frac{1}{2} \kappa \nabla_\alpha \nabla_\beta h + \bigo{\kappa^2} \eqend{,} \\
\tilde{R} &= R - \kappa h^{\alpha\beta} R_{\alpha\beta} + \kappa \nabla^\alpha \nabla^\beta h_{\alpha\beta} - \kappa \nabla^2 h + \bigo{\kappa^2} \eqend{.}
\end{equations}
Higher orders can then be obtained by repeating the expansion, i.e.,
\begin{splitequation}
F[\tilde{g}] &= F[g] + \kappa \int \left[ \frac{\delta F[\tilde{g}]}{\delta \tilde{g}_{\mu\nu}(x)} \right]_{\tilde{g} = g} h_{\mu\nu}(x) \sqrt{-g} \total^n x \\
&\quad+ \frac{1}{2} \kappa^2 \iint \left[ \frac{\delta F[\tilde{g}]}{\delta \tilde{g}_{\mu\nu}(x) \delta \tilde{g}_{\rho\sigma}(y)} \right]_{\tilde{g} = g} h_{\mu\nu}(x) h_{\rho\sigma}(y) \sqrt{-g} \total^n x \sqrt{-g} \total^n y + \bigo{\kappa^3} \eqend{,}
\end{splitequation}
and the second functional derivative is calculated by setting $g = \tilde{g}$ after performing the first one, etc.

To calculate geodesic corrections, we also need the expansion of the $n$-bein, which in symmetric gauge reads~\cite{vannieuwenhuizen1981,woodard_thesis,woodard1984}
\begin{equation}
\label{appendix_metric_bein_expansion}
e_\mu{}^a = \delta_\rho{}^a \left( \delta^\rho_\mu + \frac{1}{2} \kappa h^\rho_\mu - \frac{1}{8} \kappa^2 h_{\mu\sigma} h^{\rho\sigma} \right) + \bigo{\kappa^3} \eqend{.}
\end{equation}

\section{Geodesic expansions}
\label{appendix_geodesic}

Using the expansions~\eqref{appendix_metric_expansion} and~\eqref{appendix_metric_bein_expansion}, the boundary condition for the tangent vector to the geodesic~\eqref{calculation_prelim_geodesic_bdy2} is given to second order by
\begin{equations}
\dot \chi_0^\mu(0) &= - (x-y)^\mu \eqend{,} \\
\dot \chi_1^\mu(0) &= \frac{1}{2} h^\mu_\rho(x) (x-y)^\rho \eqend{,} \\
\dot \chi_2^\mu(0) &= - \frac{3}{8} h^{\mu\rho}(x) h_{\rho\sigma}(x) (x-y)^\sigma \eqend{,} \label{appendix_geodesic_bdy_chi2}
\end{equations}
and the boundary condition for the geodesic itself is
\begin{equation}
\chi_0^\mu(0) = x^\mu \eqend{,} \qquad \chi_1^\mu(0) = \chi_2^\mu(0) = 0 \eqend{.}
\end{equation}
Expanding the geodesic equation~\eqref{calculation_prelim_geodesic_eqn}, we obtain
\begin{equations}
\ddot \chi_1^\mu(\tau) &= \left[ \frac{1}{2} \partial^\mu h_{\alpha\beta}(\tau) - \partial_\alpha h^\mu_\beta(\tau) \right] \dot \chi_0^\alpha(\tau) \dot \chi_0^\beta(\tau) \eqend{,} \\
\begin{split}
\ddot \chi_2^\mu(\tau) &= h^{\mu\nu}(\tau) \left[ \partial_\alpha h_{\nu\beta}(\tau) - \frac{1}{2} \partial_\nu h_{\alpha\beta}(\tau) \right] \dot \chi_0^\alpha(\tau) \dot \chi_0^\beta(\tau) \\
&\quad- \left[ \partial_\alpha h^\mu_\beta(\tau) + \partial_\beta h^\mu_\alpha(\tau) - \partial^\mu h_{\alpha\beta}(\tau) \right] \dot \chi_1^\alpha(\tau) \dot \chi_0^\beta(\tau) \\
&\quad+ \left[ \frac{1}{2} \partial^\mu \partial_\nu h_{\alpha\beta}(\tau) - \partial_\alpha \partial_\nu h^\mu_\beta(\tau) \right] \chi_1^\nu(\tau) \dot \chi_0^\alpha(\tau) \dot \chi_0^\beta(\tau) \eqend{.}
\end{split}
\end{equations}
Note that at second order there are additional terms involving second derivatives of the metric (in the last line), which come from evaluating the Christoffel symbols on the first-order perturbed geodesic, and we denote
\begin{equation}
h_{\alpha\beta}(\tau) \equiv h_{\alpha\beta}(\chi_0(\tau)) = h_{\alpha\beta}(x + (y-x)\tau) \eqend{.}
\end{equation}
Integrating these differential equations with the above boundary conditions, we obtain at first order
\begin{equation}
\dot \chi_1^\mu(\tau) = \frac{1}{2} h^\mu_\rho(x) (x-y)^\rho + \int_0^\tau (x-y)^\alpha (x-y)^\beta \left[ \frac{1}{2} \partial^\mu h_{\alpha\beta}(\tau') - \partial_\alpha h_\beta^\mu(\tau') \right] \total \tau' \eqend{,}
\end{equation}
and from this, using the Cauchy formula for repeated integration~\cite{dlmf},
\begin{equation}
\chi_1^\mu(\tau) = \frac{1}{2} h^\mu_\rho(x) (x-y)^\rho \tau + \int_0^\tau (\tau - \tau') (x-y)^\alpha (x-y)^\beta \left[ \frac{1}{2} \partial^\mu h_{\alpha\beta}(\tau') - \partial_\alpha h_\beta^\mu(\tau') \right] \total \tau' \eqend{.}
\end{equation}

For the second-order corrections we can perform some simplifications since not all terms will contribute to one-loop order. Since the second-order geodesic correction to the correlation function at one-loop order contains exactly two gravitons (see Figure~\ref{fig_feynman_3}), both of these gravitons come from $\chi_2^\mu$. When taking the expectation value, massless tadpoles vanish in dimensional regularisation, and we may drop all terms containing two gravitons at the same point, such as the boundary condition~\eqref{appendix_geodesic_bdy_chi2}. Integrating once, it follows that
\begin{splitequation}
\dot \chi_2^\mu(\tau) &= \int_0^\tau \left[ \partial_\alpha h^\mu_\beta(\tau') + \partial_\beta h^\mu_\alpha(\tau') - \partial^\mu h_{\alpha\beta}(\tau') \right] (x-y)^\beta \\
&\qquad\quad\times \int_0^{\tau'} (x-y)^\gamma (x-y)^\delta \left[ \frac{1}{2} \partial^\alpha h_{\gamma\delta}(\tau'') - \partial_\gamma h_\delta^\alpha(\tau'') \right] \total \tau'' \total \tau' \\
&\quad+ \int_0^\tau \left[ \frac{1}{2} \partial^\mu \partial_\nu h_{\alpha\beta}(\tau') - \partial_\alpha \partial_\nu h^\mu_\beta(\tau') \right] (x-y)^\alpha (x-y)^\beta \\
&\qquad\quad\times \int_0^{\tau'} (\tau' - \tau'') (x-y)^\gamma (x-y)^\delta \left[ \frac{1}{2} \partial^\nu h_{\gamma\delta}(\tau'') - \partial_\gamma h_\delta^\nu(\tau'') \right] \total \tau'' \total \tau' + \text{tadpoles} \eqend{,}
\end{splitequation}
and then
\begin{equation}
\chi_2^\mu(\tau) = \int_0^\tau \dot \chi_2^\mu(\tau') \total \tau' \eqend{.}
\end{equation}

Finally, the scalar field entering the correlation function reads
\begin{splitequation}
\phi(\chi(1)) &= \phi(y) + \kappa \chi_1^\mu(1) \partial_\mu \phi(y) \\
&\quad+ \kappa^2 \chi_2^\mu(1) \partial_\mu \phi(y) + \frac{1}{2} \kappa^2 \chi_1^\mu(1) \chi_1^\nu(1) \partial_\mu \partial_\nu \phi(y) + \bigo{\kappa^3} \eqend{.}
\end{splitequation}

\section{Feynman integrals}
\label{appendix_feynman}

The general one-loop massless integral with arbitrary powers is given by~\cite{smirnov2004}
\begin{equation}
\int \frac{1}{[ k^2 - \mathi 0 ]^\alpha} \frac{1}{[ (k-p)^2 - \mathi 0 ]^\beta} \frac{\total^n k}{(2\pi)^n} = \frac{\mathi}{(4\pi)^\frac{n}{2}} \frac{\Gamma\left( \frac{n}{2} - \alpha \right) \Gamma\left( \frac{n}{2} - \beta \right) \Gamma\left( \alpha+\beta-\frac{n}{2} \right)}{\Gamma\left( \alpha \right) \Gamma\left( \beta \right) \Gamma\left( n - \alpha - \beta \right) [ p^2 - \mathi 0 ]^{\alpha+\beta-\frac{n}{2}}} \eqend{.}
\end{equation}
In particular, we have
\begin{splitequation}
\label{appendix_feynman_scalar_power}
&\int \frac{\left( p^2 \right)^{\alpha-1}}{[ k^2 - \mathi 0 ]^\alpha} \frac{1}{(k-p)^2 - \mathi 0} \frac{\total^n k}{(2\pi)^n} \\
&\quad= \frac{\Gamma\left( \frac{n}{2} - \alpha \right) \Gamma\left( \alpha+1-\frac{n}{2} \right) \Gamma(n-2)}{\Gamma\left( \frac{n}{2} - 1 \right) \Gamma\left( 2-\frac{n}{2} \right) \Gamma(\alpha) \Gamma(n-1-\alpha)} \int \frac{1}{k^2 - \mathi 0} \frac{1}{(k-p)^2 - \mathi 0} \frac{\total^n k}{(2\pi)^n} \eqend{,}
\end{splitequation}
which can also be derived in coordinate space. First one easily calculates
\begin{splitequation}
\label{appendix_feynman_x2_dalembert_p}
\left( \partial^2 \right)^k (x^2)^p &= (-4)^k \frac{\Gamma(p+1)}{\Gamma(p+1-k)} \frac{\Gamma\left( 1 - \frac{n}{2} - p + k \right)}{\Gamma\left( 1 - \frac{n}{2} - p \right)} (x^2)^{p-k} \\
&= 4^k \frac{\Gamma(p+1)}{\Gamma(p+1-k)} \frac{\Gamma\left( \frac{n}{2} + p \right)}{\Gamma\left( \frac{n}{2} + p - k \right)} (x^2)^{p-k} \eqend{,}
\end{splitequation}
using $\Gamma$ function identities~\cite{dlmf} in the second equality; note that this identity is also valid for negative $k$ by analytic continuation. From the momentum space representation for the propagator~\eqref{calculation_prelim_gravprop} it follows that
\begin{equation}
G_0(x,y) \left( \partial^{-2} \right)^{\alpha-1} G_0(x,y) = - (-1)^\alpha \iint \frac{1}{[ k^2 - \mathi 0 ]^\alpha} \frac{1}{(k-p)^2 - \mathi 0} \frac{\total^n k}{(2\pi)^n} \mathe^{\mathi p (x-y)} \frac{\total^n p}{(2\pi)^n} \eqend{,}
\end{equation}
and thus, with the explicit form of $G_0$ in coordinate space~\eqref{calculation_prelim_g0} and using twice the identity~\eqref{appendix_feynman_x2_dalembert_p}, we obtain
\begin{splitequation}
&\iint \frac{\left( p^2 \right)^{\alpha-1}}{[ k^2 - \mathi 0 ]^\alpha} \frac{1}{(k-p)^2 - \mathi 0} \frac{\total^n k}{(2\pi)^n} \mathe^{\mathi p (x-y)} \frac{\total^n p}{(2\pi)^n} = \left( \partial^2 \right)^{\alpha-1} \left[ G_0(x,y) \left( \partial^{-2} \right)^{\alpha-1} G_0(x,y) \right] \\
&\quad= - c_n^2 \frac{4^{1-\alpha} \Gamma\left( 2 - \frac{n}{2} \right)}{\Gamma\left( 1 - \frac{n}{2} + \alpha \right) \Gamma(\alpha)} \left( \partial^2 \right)^{\alpha-1} [(x-y)^2]^{\alpha-n+1} \\
&\quad= - c_n^2 \frac{\Gamma(\alpha-n+2)}{\Gamma(\alpha) \Gamma(3-n)} [(x-y)^2]^{2-n} = \frac{\Gamma(\alpha-n+2)}{\Gamma(\alpha) \Gamma(3-n)} G_0(x,y) G_0(x,y) \eqend{,}
\end{splitequation}
which for integer $\alpha$ and after performing a Fourier transform is easily seen to be equal to the result~\eqref{appendix_feynman_scalar_power}, using again $\Gamma$ function identities~\cite{dlmf}. In particular, for the cases $\alpha = 2,3$ which are needed for the calculations we obtain
\begin{equations}[appendix_feynman_scalarpowers]
\iint \frac{p^2}{[ k^2 - \mathi 0 ]^2} \frac{1}{(k-p)^2 - \mathi 0} \frac{\total^n k}{(2\pi)^n} \mathe^{\mathi p (x-y)} \frac{\total^n p}{(2\pi)^n} &= - (n-3) G_0(x,y) G_0(x,y) \eqend{,} \\
\iint \frac{p^4}{[ k^2 - \mathi 0 ]^3} \frac{1}{(k-p)^2 - \mathi 0} \frac{\total^n k}{(2\pi)^n} \mathe^{\mathi p (x-y)} \frac{\total^n p}{(2\pi)^n} &= \frac{(n-3) (n-4)}{2} G_0(x,y) G_0(x,y) \eqend{.}
\end{equations}

Since the only available external tensors are the external momentum $p^\mu$ and the metric $\eta^{\mu\nu}$, we can perform integrals with tensor factors $k^\mu$ by making a general ansatz and contracting with the external momentum or the metric, leading to scalar integrals. Those can then be performed using
\begin{equation}
2 (pk) = p^2 + k^2 - (k-p)^2 \eqend{,}
\end{equation}
and we obtain for the cases of interest
\begin{equations}[appendix_feynman_tensorfactors]
\int \frac{k^\mu}{[ k^2 - \mathi 0 ]^\alpha} \frac{1}{(k-p)^2 - \mathi 0} \frac{\total^n k}{(2\pi)^n} &= \frac{p^\mu}{2} \int \left( 1 + \frac{k^2}{p^2} \right) \frac{1}{[ k^2 - \mathi 0 ]^\alpha} \frac{1}{(k-p)^2 - \mathi 0} \frac{\total^n k}{(2\pi)^n} \eqend{,} \\
\begin{split}
\int \frac{k^\mu k^\nu}{[ k^2 - \mathi 0 ]^\alpha} \frac{1}{(k-p)^2 - \mathi 0} \frac{\total^n k}{(2\pi)^n} &= \int \left[ \frac{p^\mu p^\nu}{p^2} \left( n (p^2+k^2)^2 - 4 k^2 p^2 \right) - \eta^{\mu\nu} (p^2-k^2)^2 \right] \\
&\quad\times \frac{1}{4 (n-1) p^2} \frac{1}{[ k^2 - \mathi 0 ]^\alpha} \frac{1}{(k-p)^2 - \mathi 0} \frac{\total^n k}{(2\pi)^n} \eqend{.}
\end{split}
\end{equations}
By shifting and rescaling the integration variable, it follows that
\begin{equations}[appendix_feynman_tensorfactors2]
\int \frac{k^\mu}{[ k^2 - \mathi 0 ]^\alpha} \frac{1}{(p + \tau k)^2 - \mathi 0} \frac{\total^n k}{(2\pi)^n} &= - \frac{p^\mu}{2 \tau} \int \left( 1 + \tau^2 \frac{k^2}{p^2} \right) \frac{1}{[ k^2 - \mathi 0 ]^\alpha} \frac{1}{(p + \tau k)^2 - \mathi 0} \frac{\total^n k}{(2\pi)^n} \eqend{,} \\
\begin{split}
\int \frac{k^\mu k^\nu}{[ k^2 - \mathi 0 ]^\alpha} \frac{1}{(p + \tau k)^2 - \mathi 0} \frac{\total^n k}{(2\pi)^n} &= \int \left[ \frac{p^\mu p^\nu}{p^2} \left( n (p^2 + \tau^2 k^2)^2 - 4 \tau^2 k^2 p^2 \right) - \eta^{\mu\nu} (p^2 - \tau^2 k^2)^2 \right] \\
&\quad\times \frac{1}{4 (n-1) \tau^2 p^2} \frac{1}{[ k^2 - \mathi 0 ]^\alpha} \frac{1}{(p + \tau k)^2 - \mathi 0} \frac{\total^n k}{(2\pi)^n} \eqend{,}
\end{split}
\end{equations}
and
\begin{equations}[appendix_feynman_tensorfactors3]
\begin{split}
&\int k^\mu \frac{1}{(p + \tau k)^2 - \mathi 0} \frac{1}{((1-\tau) k - p)^2 - \mathi 0} \frac{\total^n k}{(2\pi)^n} \\
&\qquad= - p^\mu \frac{1-2\tau}{2 \tau (1-\tau)} \int \frac{1}{(p + \tau k)^2 - \mathi 0} \frac{1}{((1-\tau) k - p)^2 - \mathi 0} \frac{\total^n k}{(2\pi)^n} \eqend{,}
\end{split} \\
\begin{split}
&\int k^\mu k^\nu \frac{1}{(p + \tau k)^2 - \mathi 0} \frac{1}{((1-\tau) k - p)^2 - \mathi 0} \frac{\total^n k}{(2\pi)^n} \\
&\qquad= \left( \frac{n p^\mu p^\nu - \eta^{\mu\nu} p^2}{4 (n-1) \tau^2 (1-\tau)^2} - \frac{p^\mu p^\nu}{\tau (1-\tau)} \right) \int \frac{1}{(p + \tau k)^2 - \mathi 0} \frac{1}{((1-\tau) k - p)^2 - \mathi 0} \frac{\total^n k}{(2\pi)^n} \eqend{.}
\end{split}
\end{equations}
We also need shifted and rescaled scalar momentum integrals, which are given by
\begin{equations}[appendix_feynman_geodesic_integrals]
\int \frac{1}{(p + \tau k)^2 - \mathi 0} \frac{1}{((1-\tau) k - p)^2 - \mathi 0} \frac{\total^n k}{(2\pi)^n} &= \left( \tau (1-\tau) \right)^{2-n} \int \frac{1}{k^2 - \mathi 0} \frac{1}{(k-p)^2 - \mathi 0} \frac{\total^n k}{(2\pi)^n} \eqend{,} \\
\int \frac{1}{[ k^2 - \mathi 0 ]^\alpha} \frac{1}{(p + \tau k)^2 - \mathi 0} \frac{\total^n k}{(2\pi)^n} &= \tau^{2\alpha-n} \int \frac{1}{k^2 - \mathi 0} \frac{1}{(k-p)^2 - \mathi 0} \frac{\total^n k}{(2\pi)^n} \eqend{.}
\end{equations}

\providecommand\newblock{\ }
\bibliography{literature}

\end{document}